\documentclass{ieeeaccess}

\usepackage{cite}
\usepackage{amsmath,amssymb,amsfonts}
\usepackage{algorithmic}
\usepackage{graphicx}
\usepackage{textcomp}

\usepackage{url}
\usepackage{float}
\usepackage{nccmath}
\usepackage{adjustbox}
\usepackage{multirow}
\usepackage{eurosym}
\usepackage{mathtools}
\usepackage{textcomp}
\usepackage{verbatim}
\usepackage{soul} 
\usepackage{enumitem}
\usepackage{physics}
\usepackage{dsfont}
\usepackage{color}
\usepackage{balance}
\usepackage{subfigure}
\usepackage{nccmath}
\usepackage{algorithm}
\usepackage{array}
\usepackage{stfloats}
\newcommand{\blue}[1]{#1}

\def\BibTeX{{\rm B\kern-.05em{\sc i\kern-.025em b}\kern-.08em
    T\kern-.1667em\lower.7ex\hbox{E}\kern-.125emX}}

\begin{document}
\history{}
\doi{10.1109/TQE.2024.3398410}

\title{Variational Quantum Algorithms for the Allocation of Resources in a Cloud/Edge Architecture}
\author{
\uppercase{Carlo Mastroianni}\authorrefmark{1},
\uppercase{Francesco Plastina}\authorrefmark{2},
\uppercase{Jacopo Settino}\authorrefmark{1,2}, 
and \uppercase{Andrea Vinci}\authorrefmark{1}, 
}
\address[1]{National Research Council of Italy - Institute for High-Performance Computing and Networking (ICAR-CNR), Via P. Bucci, 8/9 C, 87036 Rende (CS), Italy
(email: carlo.mastroianni@icar.cnr.it, andrea.vinci@icar.cnr.it)}
\address[2]{Dip. Fisica, Universit\`a della Calabria, Arcavacata di Rende (CS), Italy, and INFN - Gruppo Collegato di Cosenza (e-mail: francesco.plastina@fis.unical.it, jacopo.settino@unical.it)}
%

\corresp{Corresponding author: Andrea Vinci (email: andrea.vinci@icar.cnr.it).}


\begin{abstract}

Modern Cloud/Edge architectures need to orchestrate multiple layers of heterogeneous computing nodes, including pervasive sensors/actuators, distributed Edge/Fog nodes, centralized data centers and quantum devices. The optimal assignment and scheduling of computation on the different nodes is a very difficult problem, with NP-hard complexity. In this paper, we explore the possibility of solving this problem with Variational Quantum Algorithms, which can become a viable alternative to classical algorithms in the near future. In particular, we compare the performances, in terms of success probability, of two algorithms, i.e., Quantum Approximate Optimization Algorithm (QAOA) and Variational Quantum Eigensolver (VQE). The simulation experiments, performed for a set of simple problems,
show that the VQE algorithm ensures better performances when it is equipped with appropriate circuit \textit{ansatzes} that are able to restrict the search space. Moreover, experiments executed on real quantum hardware show that the execution time, when increasing the size of the problem, grows much more slowly than the trend obtained with classical computation, which is known to be exponential.
\footnote{© 2024 IEEE. Personal use of this material is permitted. Permission from IEEE must be obtained for all other uses, in any current or future media, including reprinting/republishing this material for advertising or promotional purposes, creating new collective works, for resale or redistribution to servers or lists, or reuse of any copyrighted component of this work in other works.} 
\end{abstract}


\titlepgskip=-15pt

\maketitle

\section{Introduction}
\label{secIntro}

Cloud/Edge architectures are required to include and integrate heterogeneous types of computing devices,  with very different capabilities and characteristics, ranging from pervasive Internet of Things (IoT) sensors and actuators and mobile devices to personal computers, local servers and Cloud data centers. This complex architecture is sometimes referred to as a ``continuous computing'' platform \cite{ContinuousComputing,baresi2019unified, cogito2023,cohen2023dynamic}, and is the subject of intense research aiming to achieve a judicious allocation of computational, storage, and network resources to meet the demands of modern applications. 
%
A skillful resource allocation involves the strategic distribution of resources to the different layers, in order to deliver desired outcomes efficiently. This problem encompasses multiple dimensions, including cost-effectiveness, energy efficiency, security, and scalability, and its solution becomes increasingly intricate\cite{ResourceAllocationAdHoc2023, ResourceAllocationIEEEIndustrial}.

\blue{More specifically, the necessity emerges to combine the benefits deriving from the computation on IoT or Edge devices -- for example, real-time access to local data and easier management of privacy issues -- with the availability of more powerful remote devices, such as Cloud platforms, for the computation of intensive tasks. This necessity has created a challenge called \textit{task offloading}, which consists of determining how to partition processes among local and Cloud devices \cite{OffloadingEdgeCloud2021,JointTaskOffloading, OffloadingIoT,TaskOffloading}}.
\blue{This task can be seen as a variant of a combinatorial NP-hard problem that is well-known in the operation research community, i.e., the ``multiple knapsack problem'', where the goal is to find the best way to assign items (in our case, processes) to different knapsacks (Edge or Cloud nodes)
having different capacities \cite{MKP_models}. Assignment problems of this type can be expressed as Mixed-Integer Nonlinear Programming (MINLP) problems, for which the difficulty of obtaining closed-form solutions derives from the presence of non-convex constraints.}
\blue{Classical -- i.e., non-quantum -- multiple knapsack assignment algorithms can be categorized into three classes \cite{ MKP_review}: exact algorithms based, e.g., on branch and bound and dynamic programming techniques; approximate algorithms, which typically require a polynomial computing time and can find sub-optimal solutions with predefined error bounds, and heuristic/AI-based algorithms, including evolutionary, swarm intelligence and bio-inspired algorithms, which do not offer specific guarantees on their execution time and on their effectiveness \cite{diff_evolution}.}


A further and recent alternative is offered by the research on Quantum Computing algorithms, which, in the next years, \blue{are expected to be able to tackle combinatorial optimization problems that are intractable on classical computers, among which the task offloading/assignment problem discussed above}.
As the state of a registry of quantum bits, or qubits, can be expressed as a superposition of a number of configurations that is the exponential of the number of its qubits, the potential advantage of quantum algorithms relies on the fact that quantum gates operate in parallel on all of these configurations. 
With the Noisy Intermediate-Scale Quantum (NISQ) quantum computers available today \cite{preskillNisq, NISQ-IEEE-TQE}, the most promising approach appears the use of hybrid algorithms \cite{biamonte2017,dunjko2016, QuantumOptimization2018}, which combine classical and quantum computation and try to exploit the benefits of both.

Specifically, with Variational Quantum Algorithms (VQAs), quantum computation is driven by a set of parameters that are optimized classically, in a cycle that aims at finding the best solution with a significant speed-up with respect to classical approaches \cite{VQA, VQA-theory, BarrenPlateaus2023}. On the one hand, VQAs facilitate the design of task-oriented quantum programs: while pure quantum algorithms can be non-intuitive and difficult to adapt to most problems, it is often possible to reformulate the latter in a format that is addressable for VQAs, which is, at least in part, a standard procedure. On the other hand, VQAs are presumed to need small qubit counts and shallow quantum circuits, since the optimization of parameters is entrusted to classical computational algorithms.


In this paper, following some preliminary ideas sketched out in \cite{CF2023}, we discuss how a typical resource management problem can be reformulated in order to be solved by two renowned Variational Quantum Algorithms, i.e., the Quantum Approximate Optimization Algorithm (QAOA) \cite{QAOA}, and the Variational Quantum Eigensolver (VQE) \cite{VQE}. In particular, we address the problem of assigning a number of processes to a set of Edge nodes or to the Cloud. An efficient assignment must achieve a compromise between the advantages offered by Edge nodes -- lower latency, more efficient usage of local data, higher level of security, etc. -- and the larger computing and memory capabilities ensured by Cloud facilities.

The optimization problem can be mapped
into an Ising problem \cite{ising2000}, and then solved by QAOA and VQE. In particular, a quantum ``Hamiltonian'' operator is built from the Ising expression, and, then,  the aim becomes that of finding the minimum eigenvalue of this operator, together with the corresponding eigenvector, i.e., the ``ground state'', in the Physics language. As we will show in this work, the components of this eigenvector correspond to the values of the binary variables that determine the assignment of the processes to an Edge node or to the Cloud.

\blue{This paper aims to contribute to the current research in the following ways:
\begin{enumerate}
    \item we provide the assessment and comparison between the performances of the QAOA and VQE algorithms. While the theory predicts that QAOA converges to the optimal solution when increasing the number of gate repetitions, the results show that the improvement is slow, not only in the presence of noise, whose impact depends on the circuit depth, but also in noiseless simulations, due to the larger number of parameters that must be optimized. We have found that VQE over-performs QAOA thanks to the definition of ad-hoc circuit ansatzes, motivated by the specific problem, which can reduce the size of the search space and provide shortcuts toward the optimal solution;
    \item after concluding about the superiority of VQE, we compare the performance of four different VQE ansatzes. The results show a performance improvement when the ansatz is  better tailored to the specific problem, and, in particular, when the correlations among variables, induced by the problem, are matched by the gates of the quantum ansatz;
    \item for a more complete analysis, we consider some variants of the optimization problem.
    An important design choice is whether an Edge node can be loaded up to its full capacity or only for a certain fraction. This choice originates from the large amount of literature demonstrating that the computing resources are better exploited, and the energy consumption is lower, if the nodes are either loaded at a high fraction of their capacity or hibernated. Therefore, it can be convenient to consolidate the load on fewer but heavily loaded nodes \cite{WSN_SleepScheduling,WSN_DutyCycling};
    \item we report results in terms of scalability, when increasing the problem size. Though the current size and quality of quantum hardware do not permit a definite answer, the different trends of classical and quantum algorithms in the increase of computing time support the expectations that, in the future, bigger and more efficient quantum computers will be able to provide a real speed-up.
\end{enumerate}
}

\blue{As a final comment, we are aware that in recent years there has been large amount of work, and many tentative answers, regarding the possible achievement of a quantum advantage for optimization and machine learning problems. In \cite{BetterThanClassical}, the authors report a deep investigation on this aspect, where the conclusion is that no speed-up can be proven so far, but also that there is room for more research, especially in the attempt to adapt the quantum algorithms to specific problems and input data. Along this avenue, our main intent is to help the readers discriminate among the possible choices for tackling an important problem, using the Edge/Cloud scenario as a significant and concrete application domain, and, in this way, hopefully accelerate the quest for quantum advantage.}

%

The paper is organized as follows: Section \ref{secRelated} discusses related works in this field;
Section \ref{secModeling} illustrates the reference Edge/Cloud architecture, introduces the assignment problem in this context, and shows how the  problem can be transformed into an Ising problem, suitable for quantum computation; Section \ref{secQuantumAlgorithmsQaoaVqe} illustrates the QAOA and VQE algorithms and the related quantum circuits and, for VQE, reports four different variants, or ``ansatzes''; Section \ref{secResults} reports a wide set of performance results, in terms of success probability and execution time. Finally, Section \ref{secConclusions} concludes the paper and discusses some avenues for future research work.














\section{Related work}
\label{secRelated}



The availability of a huge number of Internet-of-Things devices, and their pronounced heterogeneity, has given the opportunity and the necessity of integrating these devices with pre-existent centralized infrastructures, such as data centers and cloud platforms, thus converging towards the ``continuous computing'' paradigm \cite{ContinuousComputing,baresi2019unified, cogito2023}. The computing power and flexibility ensured by this paradigm has favored the birth, or the significant improvement, of 
a wide plethora of applications, including infotainment, road
safety and virtual network functions \cite{cohen2023dynamic}. For example, a road safety application requires low latency, and can be processed by mobile devices with limited capabilities, e.g., in terms of computing power and battery duration, which need to be supported by edge servers, close to the users, and/or by centralized data centers \cite{zhao2019mobile}.
%
%
Because of this, several key challenges emerged in this realm, among which: (i) the need for protocols and standards that enable the manipulation of heterogeneous devices and infrastructures; (ii) the support of security and privacy requirements, and (iii) the design of efficient resource allocation and scheduling algorithms that aim at optimizing the exploitation of the available resources, while addressing their capacity constraints \cite{raeisi2023resource, RM-survey}.


Recent research in this area is connected to machine learning and AI-driven decision-making algorithms.
Indeed, efficiency, latency, and privacy issues foster the deployment of machine learning algorithms towards the edge of the network \cite{EdgeComputingDeepLearning2020}.
At the same time, AI-driven approaches can be used to control the offloading of applications
\cite{OffloadingEdgeCloud2021}. Though many heuristics have been proposed, when the size of the problem increases their exploitation is difficult and does not offer performance guarantees. A recent strategy is to exploit reinforcement and deep reinforcement learning techniques that learn to make decisions based on the feedback (reward) received from the users and the environment \cite{RL-liu2019resource, DRL-chen2018optimized}. A promising approach is to adopt the federated learning paradigm, where edge devices train the models using local data, and send the learned parameters to a server, which builds an aggregate model and sends it back to the edge layer, in a cycle that ends when the loss function reaches convergence \cite{FL-wang2019adaptive, FL-abreha2022federated}.

Quantum computing algorithms promise to be a viable alternative for the solution of optimization problems \cite{VQA, VQA-IEEE-TQE}, as, for example, Max-Cut \cite{maxcut}, Max-Sat \cite{maxsat} and routing problems \cite{routing-IEEE-TQE}. In \cite{IsingNP}, hints are given on how an NP-problem can be formulated as an Ising problem and then solved with the adiabatic paradigm, where a quantum system is driven towards the solution, expressed as the state of minimal energy of its Hamiltonian. This procedure has inspired  algorithms that can be run on a gate-based quantum computer, such as those provided by major IT companies. As an example, we employed such an approach in a previous work on the optimization of energy exchanges within a prosumer community \cite{IEEESmartGrid}.
%
%
The applications of quantum algorithms to edge computing are all very recent. In \cite{ResourceMgtEdge}, a quantum-inspired reinforcement learning (QRL) technique is proposed to determine the best strategy to offload part of the computation from mobile devices to edge servers. The authors of \cite{HybridNetOptimization} describe how the network resource allocation can be determined by expressing a MILP/MINLP problem with the QUBO model. In \cite{qu2022secure}, the focus is on security issues, and a new quantum fog computing model is proposed to ensure a high level of protection and thwart a variety of fog computing attacks.

Besides using standard procedures for classical-to-quantum problem conversion, it can be very useful to adapt the quantum algorithm to the specific problem, as this can boost efficiency and increase the success probabilities.
Indeed, when using Variational Quantum Algorithms, the peculiarities of a problem can be reflected in the definition of the ansatzes, i.e., the parameterized quantum circuits that prepare the quantum states. As opposed to ``hardware-efficient'' ansatzes, which aim to reduce the depth and complexity of the quantum circuit \cite{hardware-efficient}, ``problem-inspired'' ansatzes also use information about the problem in order to tailor a more focused ansatz.

QAOA \cite{QAOA} and VQE \cite{VQE} offer two different approaches to ``bring the problem into the ansatz''. In QAOA, inspired by the adiabatic approach, two different quantum Hamiltonians (the first of which expresses the problem) take turn in driving the evolution towards the state of minimal energy of the first, which encodes the solution of the optimization problem. The main advantage of QAOA is that it is known to converge when increasing the number of repetitions of the quantum circuit. Two major drawbacks, however, are: (i) a large number of repetitions, and therefore a large circuit depth, make the algorithm highly prone to noise and decoherence and (ii) QAOA does not try to reduce the search space, which grows exponentially with problem size. On the other hand, VQE leaves to the designer the possibility, but also the effort, of tailoring the ansatz for the specific problem, but at the cost of losing any convergence guarantee. In particular, our proposal is to build a number of ansatzes that prepare a proper superposition of Dicke states -- see \cite{Dicke1-IEEETQC, Dicke2-IEEETQC} -- with Hamming weight equal to 1, i.e., in which only one qubit at a time takes the state $\ket{1}$.

\section{Modelling of the Edge/Cloud Assignment Problem}
\label{secModeling}

This section includes three subsections that focus, respectively, on: (A) the reference quantum-assisted Edge/Cloud architecture on which we focus; (B) the formulation of the assignment problem as a linear programming problem, in four different versions; and (c) the steps needed to translate the original problem into an Ising one, which can be tackled with Variational Quantum Algorithms such as QAOA and VQE.

\subsection{Reference Architecture}

In the next few years, the research and industrial developments will most probably allow quantum facilities to be integrated into the current continuous computing paradigm
\cite{ContinuousComputing,baresi2019unified, cogito2023,cohen2023dynamic}, resulting in a novel and more advanced architecture.
%
%
In such a composite architecture, resource management algorithms need to be used to determine where the computation should be performed (assignment problem) and when (scheduling problem)\cite{amadeo2023network}.


\blue{Figure \ref{fig:architecture} shows a sketch of the quantum-assisted Edge/Cloud computing continuum architecture exploited as a reference in this work. This architecture has become a standard in recent years -- see \cite{OffloadingEdgeCloud2021,FutureEdgeCloud, JointTaskOffloading}  -- and, starting from it, we propose the addition of quantum resources, whose role will be discussed below in this section.}
%
%
\blue{At the bottom, the Device layer includes a set of IoT end-devices, which span from connected sensors/actuators to smart objects and users' smartphones. An IoT device can be connected to a specific Edge node or can access the Internet directly.} Devices can gather sensed information, act upon an environment, do local processing, and request processing to the Edge/Cloud continuum. At this level, resource management algorithms are used, for example, to build wireless sensor networks that are able to minimize latency and energy consumption\cite{wsn}.

\Figure[!tb]()[width=0.9\linewidth]{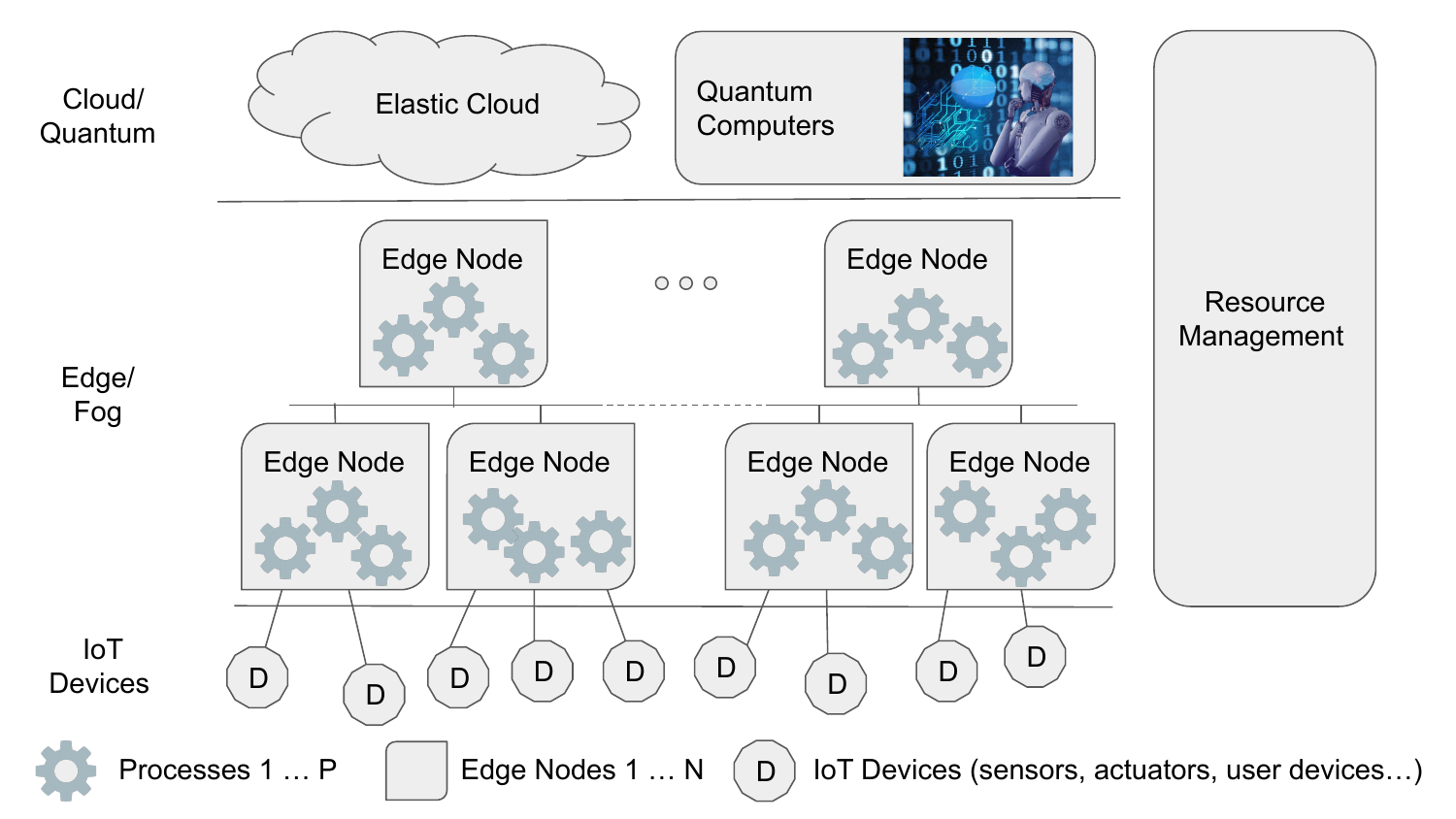}
{Sketch of Quantum-assisted Edge/Cloud computing continuum.
\label{fig:architecture}}

The intermediate Edge/Fog layer consists of a network of heterogeneous computing nodes available at different levels of the networking infrastructure that connects the device layer and the internet. These nodes can be deployed near the devices, e.g., in-home Edge computing nodes, or made available by network providers, e.g., in the nearing of cellular radio towers. The computational capabilities and resources may vary widely; however, in general, the nearer to the end devices are the nodes, the lower their capabilities. But, on the other side, processing on Edge nodes results in lowering the network latency in communication with the end devices. Within this layer, resource management algorithms are used for several purposes; for example, to select the Edge node(s) to which the devices should be connected, and to determine which processes should be offloaded to the device layer or to the Cloud \cite{CloudOfThings}

The top layer includes and integrates Cloud and quantum resources. They are positioned at the same level as they need to be tightly integrated with each other, since hybrid algorithms combine classical and quantum computation, as mentioned in the introductory section.
Adding quantum resources to the typical Cloud/Edge architecture can help to solve hard problems for which quantum algorithms promise to offer a computational speedup in the coming years; specifically, optimization problems and machine learning problems.
Currently, resource management algorithms are used in the Cloud layer to determine the assignment and scheduling of processes on the nodes of a Cloud data center \cite{SICloud, altomare2019data}. With the inclusion of quantum platforms, two novel research avenues have been opened in this context: (i) on the one hand, resource management algorithms are needed to assign and schedule specific parts of the computation on quantum hardware; (ii) on the other hand, there are chances that such algorithms, applied to any layer of the architecture, can be executed more efficiently on quantum hardware than on classical computers. In some cases, when problem instances are dimensionally large, and classical computation is infeasible, quantum computers can even become the only available opportunity to tackle the given task. In this paper, we are interested in the latter type of problems, and investigate how a quantum algorithm can be used to partition resources between a set of Edge nodes and the Cloud, as detailed in the following.


\subsection{Formulation of the Assignment Problem}
\label{SubSecProblems}

In particular, we cope with four different versions of the Edge/Cloud assignment problem, differentiated by the following alternatives:

\begin{enumerate}
    \item the first alternative consists of letting a process be assigned only to an Edge node, or also to the Cloud. The first case occurs either when the Cloud is not present, or when the set of processes that need to be assigned to the Edge layer has been determined previously, for example based on privacy issues;
    \item the second alternative regards the load that can be assigned to an Edge node, i.e., whether it can be loaded at any fraction of its capacity, or at a predefined minimum capacity, for example 70\% or 80\%. The latter choice is frequently adopted to reduce the energy consumption. The rationale is that an active but lowly loaded node consumes a notable amount of power, whereas an unloaded node consumes a negligible amount of power if it is put in a sleep mode. Therefore, it is convenient to consolidate the load on fewer but heavily loaded nodes, and put the other nodes in sleep mode, see Refs: \cite{WSN_SleepScheduling} and \cite{WSN_DutyCycling}.
\end{enumerate}

Based on these two choices, we will refer to the four problems with the following names and acronyms: ``Edge and Cloud, Free Load'' (ECFL), ``Edge Only, Free Load'' (EOFL), ``Edge and Cloud, High Load'' (ECHL), and ``Edge Only, High Load'' (EOHL). In the following, we first describe the most general problem, which is the ECFL one, then we specify the modifications that apply in the other cases. For each problem, we determine the size of the problem, i.e., the number of binary variables $Q$, which, in a quantum circuit, becomes the number of qubits needed to solve the problem.

\textbf{ECFL problem.} Given a set of processes $\mathcal{P}=\{1, ..., P\}$ and a set of computational Edge nodes $\mathcal{N}=\{1, ..., N\}$, the problem is to decide whether to assign each process to the Edge layer (and, in this case, decide to which node) or to the Cloud.
The execution of processes on Edge nodes can bring important benefits, for example, in terms of service latency and access to local data that have peculiar security and privacy characteristics. The assignment must combine these benefits with the limited capacity of Edge nodes.



\blue{As mentioned in the introductory section, the problem can be formulated as a binary linear programming problem, or more specifically, as a multiple knapsack problem, which is known to be NP-hard~\cite{conforti2014integer}.} A set of binary variables $x_{ij}$ are defined, which take the value $1$ if the process $i$ is assigned to the Edge node $j$, and $0$ otherwise. If a process $i$ is not assigned to any Edge node, it is assigned to the Cloud. The objective is to assign the processes to nodes by maximizing an overall gain function, while matching the capacity constraints of the Edge nodes. The gain function and the constraints can be written:

\begin{align}
     \label{eq:objFunction}
     \max & \sum_{ i \in \mathcal{P},j \in \mathcal{N}} v_{ij}x_{ij}  \\
     \label{eq:vincoliProcesso}
    & \sum_{j \in \mathcal{N}} x_{ij} \leq 1 , ~~~ & \forall i \in \mathcal{P} \\ 
    \label{eq:vincoliNodo}
    & \sum_{i \in \mathcal{P}} w_{i}x_{ij} \leq B_{j} ,~~~ & \forall j \in \mathcal{N} 
\end{align}
where each process $i \in \mathcal{P}$ is assigned a value $v_{ij}$, which is the value gain of executing process $i$ on a node $j \in \mathcal{N}$, w.r.t. executing the same process on the Cloud (e.g., the value can be a measure of effectiveness in terms of latency or occupied network bandwidth), and an integer weight $w_i$, which represents the amount of computing resources required by process $i$ for its execution. Each Edge node $j$ has a capacity $B_j$, defined as an integer.
It is assumed that the Cloud capacity is much larger than the capacity of the Edge nodes, therefore no constraint is defined on the Cloud load.



The number of binary variables $x_{ij}$ is equal to $P \cdot N$. However, the reformulation into an Ising problem, later discussed in Section \ref{subsecStepsToIsing}, needs the transformation of inequalities into equations. In order to perform this transformation, a number of slack binary variables must be added. In particular, for each process $i$, one slack binary variable $p_i$ is needed to specify whether the process is actually assigned to an Edge node or to the Cloud. Accordingly, by adding $P$ slack variables, constraints (\ref{eq:vincoliProcesso}) are reformulated as:

\begin{equation}
\label{eq:residualP}
\sum_{j \in \mathcal{N}} x_{ij} + p_i = 1
\end{equation}

Furthermore, to convert the inequalities (\ref{eq:vincoliNodo}) into equations, for each node $j$, we need a set of slack binary variables, which allow giving every possible value to the residual capacity of the node $j$, i.e., the capacity that remains not assigned to any process, defined as:

\begin{equation}
\label{eq:residualB}
     r_j = B_j - \sum_{i \in \mathcal{P}} w_{i}x_{ij}
\end{equation}

Since the node can be filled at any fraction of its capacity, $r_j$ can assume any value between 0 and $B_j$, and the number of slack binary variables is $\lceil \log_2 (B_j+1) \rceil$. Using these slack variables, which are denoted as $b_{jk}$, the constraints (\ref{eq:vincoliNodo}) can be formulated as:

\begin{align}
\label{eq:ineq_to_eq}
    & \sum_{i \in \mathcal{P}} w_ix_{ij} + \sum_{k=1}^{\lceil \log_2(B_j+1) \rceil} 2^{(k-1)} b_{jk} = B_j
\end{align}

The overall number of binary variables is then:

\begin{equation}
\label{eq:nQubits}
    Q = P \cdot (N + 1) + 
    \sum_{j \in \mathcal{N}}
    \lceil \log_2(B_j+1) \rceil
\end{equation}


%


\textbf{EOFL problem.} In this case, the processes can only be assigned to one of the Edge nodes. Accordingly, the only difference is that the inequality constraints (\ref{eq:vincoliProcesso}) become equations:

\begin{equation}
\sum_{j \in \mathcal{N}} x_{ij} = 1
\end{equation}
\\
Slack variables are not needed for these constraints, while they are still necessary for the inequalities (\ref{eq:vincoliNodo}). Overall, the number of binary variables is:

\begin{equation}
Q = P \cdot N +  \sum_{j \in \mathcal{N}}
    \lceil \log_2(B_j+1) \rceil    
\end{equation}

\textbf{ECHL problem.} For this problem, the assumption is made that the load of each Edge node $j$ must be equal or greater than a specified threshold $T_j$, with $0 < T_j < B_j$. 
The constraints (\ref{eq:vincoliNodo}) become:

\begin{equation}
\label{eq:threshold}
T_j \leq \sum_{i \in \mathcal{P}} w_{i}x_{ij} \leq B_{j}
\end{equation}
\\
The residual capacity $r_j$, defined in (\ref{eq:residualB}), must lie between 0 and the quantity defined as:

\begin{equation}
\label{eq:residualB2}
\hat{B}_j = B_j - T_j
\end{equation}

This is obtained from (\ref{eq:threshold}), by multiplying each member by $-1$ and then adding the quantity $B_j$. Hence, the number of slack variables needed for each node is equal to $\lceil \log_2 (\hat{B}_j+1) \rceil$, which,  in general, is lower than $\lceil \log_2 (B_j+1) \rceil$. Using these slack variables, the constraints (\ref{eq:vincoliNodo}) are formulated as:

\begin{align}
\label{eq:ineq_to_eq_hat}
    & \sum_{i \in \mathcal{P}} w_ix_{ij} + \sum_{k=1}^{\lceil \log_2(\hat{B}_j+1) \rceil} 2^{(k-1)} b_{jk} = B_j
\end{align}

%
For the ECHL problem, the number of binary variables is then:

\begin{equation}
Q = P \cdot (N+1) +  \sum_{j \in \mathcal{N}}
    \lceil \log_2(\hat{B}_j+1) \rceil    
\end{equation}


\textbf{EOHL problem.} Following the above considerations, this problem is defined as: 

\begin{align}
     \max & \sum_{ i \in \mathcal{P},j \in \mathcal{N}} v_{ij}x_{ij}  \\
    & \sum_{j \in \mathcal{N}} x_{ij} = 1 , ~~~ & \forall i \in \mathcal{P} \\ 
    & T_j \leq \sum_{i \in \mathcal{P}} w_{i}x_{ij} \leq B_{j} ,~~~ & \forall j \in \mathcal{N} 
\end{align}
\\
The number of binary variables is:

\begin{equation}
Q = P \cdot N +  \sum_{j \in \mathcal{N}}
    \lceil \log_2(\hat{B}_j+1) \rceil    
\label{eq:Q_EOHL}
\end{equation}


\subsection{Transformation to an Ising Problem}
\label{subsecStepsToIsing}

The optimization problem shown in the previous section can be tackled with a gate-based quantum computing. This requires the preliminary reformulation of the ILP problem into an Ising problem. 
Such a reformulation is described below, with reference to the ECFL problem.  After converting the inequality constraints into equations, as anticipated in Section \ref{SubSecProblems}, the next step is to include the constraints (\ref{eq:residualP}) and (\ref{eq:ineq_to_eq}) into the objective function (\ref{eq:objFunction}), in the form of penalties. An extended objective function is defined as:

\begin{multline} \label{eq:extObj}
     \min \bigg( - \sum_{i \in \mathcal{P},j \in \mathcal{N}}v_{ij}x_{ij}  + \\
      A \cdot \sum_{j \in \mathcal{N}}\Big(
        B_j - \sum_{i \in \mathcal{P}} w_ix_{ij} + \sum_{k=1}^{\lceil \log_2(B_j+1) \rceil} 2^{(k-1)} b_{jk}
      \Big)^2 + \\
      A \cdot \sum_{i \in \mathcal{P}}\Big(
      1 - \sum_{j \in \mathcal{N}} x_{ij} + p_i\Big)^2     
      \bigg)
\end{multline}

where the value of the constant $A$ is defined as:

\begin{equation}
A = 1 + \sum_{ i \in \mathcal{P},j \in \mathcal{N}}v_{ij}
\label{eq:penalty}
\end{equation}

The maximization problem has been converted to a minimization problem, and each constraint has been transformed into a penalty, which is equal to 0 only when the constraint is satisfied by the values of the binary variables. When the constraint is not satisfied, the value of the penalty is equal to or larger than $A$. Since the value of $A$ is defined to be larger than the maximum possible value of the first term of Eq. (\ref{eq:extObj}),
the violation of even a single constraint cannot be compensated by the minimization of the first term. This ensures that the solution that minimizes the objective function is obtained with values of the binary variables that satisfy all the constraints.



The following step is to replace the $Q$ binary variables with corresponding discrete variables. In particular, for every binary variable $\tilde{x}_i, i=1 \cdots Q$, with $i$ globally running on all the binary (including slack) variables, a corresponding discrete variable $z_i$ is defined with the substitution:

\begin{equation}
\label{eq:substitutionx}    
   \tilde{x}_i =\frac{1-z_i}{2}, ~~~ i = 1, ..., Q 
\end{equation}

After the substitution, the extended objective function is rewritten as a sum of terms $z_i$ and $z_i \cdot z_j$, and an Ising problem is obtained \cite{ising2000}, formulated as:

\begin{equation}
\label{eq:ising}
	min \ \bigg( {\sum_{i=1}^{Q}{h_i \cdot z_i} - \sum_{i=1}^{Q}\sum_{j=1}^{i-1}{J_{ij} \cdot z_i \cdot z_j}} \bigg)
\end{equation}
\\
where $h_i$ and $J_{ij}$ are real constants obtained after applying the substitutions.
\blue{Details concerning the correspondence between the coefficients of the Ising problem and the parameters of the ILP problem are provided in the Appendix.}

Now that we have obtained an Ising problem, we can approach it with a quantum computing algorithm, as discussed in detail in the following section.

%


\section{Quantum Algorithms: QAOA and VQE}
\label{secQuantumAlgorithmsQaoaVqe}

As mentioned in the introductory section, we leverage Variational Quantum Algorithms (VQAs) for solving the Edge/Cloud assignment problem. 
In order for this paper to be more self-contained, this section succinctly reiterates the primary objectives of these algorithms and delineates their specific application within our 
framework.
Thus, we summarize here the main features of two specific algorithms, QAOA and VQE, and, for the latter one, we describe four different ansatz formulations that will be compared in the results section when applied to our specific problem.

Given the correspondence with the Ising-like Hamiltonian carried out in Section \ref{subsecStepsToIsing}, the quest for its ground state coincides with that for the best solution of the optimization problem. The latter being defined in terms of a set of $Q$ discrete binary variables $\{z_i\}$, taking the values $+1$ and $-1$, its solution is one of the $2^Q$ possible strings of these values.

In order to employ a quantum approach, each variable is associated with one of the qubits of a quantum register. In particular, $z_i$ is given by the outcome of the measurement of one of the Pauli operators, the so-called $\textbf{Z}$ observable, performed on the $i$-th qubit at the end of the algorithm. According to quantum mechanics, the measurement has, indeed, the two possible outcomes $+1$ and $-1$, which are the two eigenvalues of  $\textbf{Z}$. Correspondingly, after the measurement, the state of each qubit collapses into one of the two logical states, denoted (using Dirac notation) by $\ket 0 = [1,0]^T$ and $\ket 1= [0,1]^T$. These are the eigenstates of the $\textbf{Z}$ operator, which
can expressed, in the logical basis, as the third Pauli Matrix:

\[
\textbf{Z} = \begin{bmatrix}
    1  & 0  \\
    0  & -1 \\
\end{bmatrix}
\]

The Ising problem (\ref{eq:ising}) is, then, mapped to a diagonal Hamiltonian operator, built with sums and tensor products (i.e., Kronecker products) of two basic one-qubit operators, the identity \textbf{I} and the Pauli operator \textbf{Z}. For each term in (\ref{eq:ising}), the operator $\textbf{Z}_i$ substitutes the variable $z_i$, and the identity operator $\textbf{I}_i$ is assumed to be inserted for each variable $z_i$ that does not appear explicitly. Moreover, the multiplications between two $z$ variables are substituted by the tensor products between the corresponding \textbf{Z} operators. For example, with $Q=4$, the term $z_2$$\cdot$$z_3$becomes $\textbf{I}_1 \otimes \textbf{Z}_2 \otimes \textbf{Z}_3 \otimes \textbf{I}_4$ or, more succinctly, $\textbf{I}_1 \textbf{Z}_2 \textbf{Z}_3 \textbf{I}_4$ or, even more briefly, $\textbf{Z}_2 \textbf{Z}_3$, where the identity operators are implicit.  With these rules, the Hamiltonian operator that corresponds to expression (\ref{eq:ising}) is:

\begin{equation}
\label{eq:IsingHamiltonian}
	\textbf{H} = {\sum_{i=1}^{Q}{h_i \cdot \textbf{Z}_i} - \sum_{i=1}^{Q}\sum_{j=1}^{i-1}{J_{ij} \cdot \textbf{Z}_i \otimes \textbf{Z}_j}}
\end{equation}

Now, the problem is to find the minimum eigenvalue(s) of the operator (\ref{eq:IsingHamiltonian}), which corresponds to finding the string of values of $z$ variables that minimizes the Ising expression (\ref{eq:ising}). 

Both the QAOA and VQE algorithms can be used to explore the Hilbert space and search for the ground state of the Ising Hamiltonian operator\footnote{VQE can be exploited also for Hamiltonian operators that include $\textbf{X}$ and $\textbf{Y}$ observables, which, however, are not present in the Hamiltonian derived from the assignment problem.}.
\subsection{QAOA}

The QAOA algorithm has been proposed in \cite{QAOA} as the reformulation, in terms of quantum gates, of an adiabatic process that, starting from the known ground state of a simple Hamiltonian, evolves towards the unknown ground state of a Hamiltonian that defines the specific optimization problem. The QAOA circuit is inspired by the Trotter approximation of the Hamiltonian time evolution.
For example, let us assume that the Ising Hamiltonian of the problem is:

\begin{multline}
\label{eq:Hamiltonian-ex}
	\textbf{H} = 
 (\textbf{Z}_1 \otimes \textbf{I}_2 \otimes \textbf{I}_3) 
 +2\ (\textbf{I}_1 \otimes \textbf{I}_2 \otimes \textbf{Z}_3)  \\
 -4\ (\textbf{Z}_1 \otimes \textbf{Z}_2 \otimes \textbf{I}_3) 
 -2\ (\textbf{I}_1 \otimes \textbf{Z}_2 \otimes \textbf{Z}_3) 
\end{multline}

\begin{figure}[!tb]
	\centering
	\includegraphics[width=0.95\columnwidth, trim=0cm 0cm 0cm 0cm]{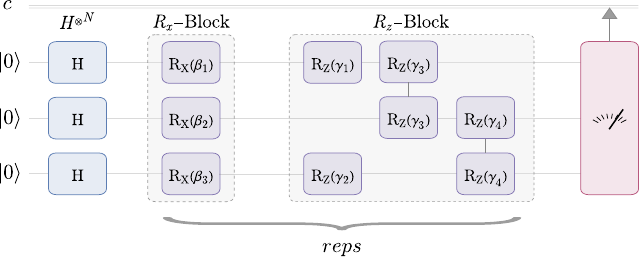}
    \caption{Example of a QAOA circuit. \blue{The circuit prepares the uniform superposition state with a set of Hadamard gates. Then, it applies, for a number of times \textit{reps}, the initial (or mixing) Hamiltonian $\textbf{H}_m = \sum_{i=1}^Q X_i$, and the problem Hamiltonian $\textbf{H}$ defined in (\ref{eq:Hamiltonian-ex}). In particular, the two single-qubit and the two double-qubit rotation gates correspond to the four terms in (\ref{eq:Hamiltonian-ex}). Please see reference \cite{QAOA} for more details on the QAOA circuit.} The objective is to tune the values of the parameters $\beta$ and $\gamma$ and prepare, before measurement, the eigenstate that corresponds to the minimum eigenvalue of the problem Hamiltonian.}
	\label{fig:QAOAcomplexity}
\end{figure}

The QAOA quantum circuit used to find the ground state of this Hamiltonian is depicted in Figure \ref{fig:QAOAcomplexity}, which shows three main groups of quantum gates, where the last two blocks are repeated a number of times, referred to as \textit{reps}.

The most relevant advantage of QAOA is that it is known to converge to the best solution when increasing the number of repetitions. However, in practical terms, this advantage cannot be fully exploited in the NISQ era, since the impact of noise increases with the circuit depth. Moreover, the number of parameters to be tuned also increases with the number of repetitions, which complicates the task of the optimization algorithm. Both these limitations originate from the fact that the search is extended to the whole Hilbert space, without any restriction that can be derived from the constraints of the problem. 

\subsection{VQE}
\label{subsecVQE}

As opposed to QAOA, with the VQE algorithm we can define a quantum circuit, or \textit{anstatz}, whose expressibility limits the boundary of the search space. This may facilitate the task of the optimization algorithm, provided we are guaranteed that the optimal solution can be found within these boundaries. 

For the Cloud/Edge problem we define four alternative ansatzes.
%
To make the discussion of the ansatzes clearer, we focus on a specific scenario, i.e., an ECFL problem -- see Section \ref{SubSecProblems} -- where three processes need to be assigned to one of two Edge nodes or to the Cloud, i.e., $P=3$ and $N=2$. The capacities of the nodes are assumed to be $B_1=3$ and $B_2=2$, while the weights of the processes are set as: $w_1=2$, $w_2=1$ and $w_3=1$. The needed number of qubits is $Q=13$, see (\ref{eq:nQubits}).
For each ansatz, we will calculate the number of parameters that need to be optimized, the number of two-qubit gates executed in the circuit, and the circuit depth in terms of two-qubit gates\footnote{We focus on two-qubit gates because their execution times and error rates are orders of magnitude bigger than those of single-qubit gates.}. These parameters will be denoted, respectively, as $\Theta$, $G_2$ and $D_2$.
 

\subsubsection{Ansatz A1}

\begin{figure}[!tb]
    \centering
    \includegraphics[width=0.90\columnwidth, height=1.15\columnwidth, trim=0cm 0cm 0cm 0cm]    {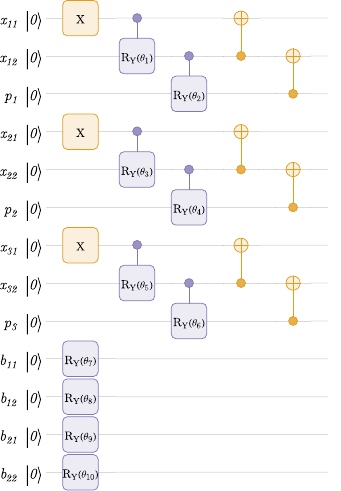}
    \caption{Ansatz A1. This ansatz prepares, for each process, a superposition of the basis states that fulfill the assignment constraint: the process is assigned to exactly one node.}
    \label{fig:ansatz_circuit}
\end{figure}

The first ansatz, or Ansatz A1, which is also the basis for the other ones, exploits the constraint that each process must be assigned to only one node, where here for ``node'' we intend either a true Edge node or the Cloud. For the scenario specified above, the ansatz is shown in Figure \ref{fig:ansatz_circuit}. The circuit uses four standard gates: two single-qubit gates, i.e., the NOT gate and the $R_Y$ rotation gate (for which the angle of rotation is a variational parameter $\theta$), and two two-qubit controlled gates, i.e, the $CNOT$ gate and the $CR_{Y}$ gate, which apply, respectively, the NOT and the $R_Y$ gate to the ``target'' qubit when the state of the ``control'' qubit is $\ket{1}$. 

The objective is to prepare, for each process, a state that is the most generic superposition of the solutions that match the mentioned constraints. For example, for the first node, an admissible solution is one that prepares exactly one of the qubits $x_{11}$, $x_{12}$, and $p_1$ to $\ket{1}$, and the other two to $\ket{0}$. The gates on the first three qubits ensure this. Indeed, when applying the gates in the sequence shown in Figure \ref{fig:ansatz_circuit}, the first three qubits take the following states:

{\footnotesize
\begin{align*}
\begin{aligned}
&\ket{000} \xrightarrow{\text{X}} \ket{100}  \xrightarrow{\text{CR}_Y(\theta_1)}\\
&\ket{1} \big(\cos\frac{\theta_1}{2}\ket{0}+\sin\frac{\theta_1}{2}\ket{1} \big) \ket{0} = \cos\frac{\theta_1}{2}\ket{100} + \sin\frac{\theta_1}{2}\ket{110} \xrightarrow{\text{CR}_Y(\theta_2)}\\
&\cos\frac{\theta_1}{2}\ket{100} + \sin\frac{\theta_1}{2}\ket{11}
\big( \cos\frac{\theta_2}{2}\ket{0} + \sin\frac{\theta_2}{2}\ket{1} \big) =^{ \textcolor{white}{S}}\\
&\cos\frac{\theta_1}{2}\ket{100} + \sin\frac{\theta_1}{2}\cos\frac{\theta_2}{2}\ket{110} + \sin\frac{\theta_1}{2}\sin\frac{\theta_2}{2}\ket{111} \xrightarrow{\text{CNOT}_{2,1}}\\
&\cos\frac{\theta_1}{2}\ket{100} + \sin\frac{\theta_1}{2}\cos\frac{\theta_2}{2}\ket{010} + \sin\frac{\theta_1}{2}\sin\frac{\theta_2}{2}\ket{011} \xrightarrow{\text{CNOT}_{3,2}}\\
&\cos\frac{\theta_1}{2}\ket{100} + \sin\frac{\theta_1}{2}\cos\frac{\theta_2}{2}\ket{010} + \sin\frac{\theta_1}{2}\sin\frac{\theta_2}{2}\ket{001}
\end{aligned}
\end{align*}
}



In the expressions above, the arrows indicate the gate executions, and, in the gates CNOT$_{i,j}$, the subscripts $i$ and $j$ are, respectively, the positions of the control qubit and the target qubit in the circuit, starting from the top. The final state is a superposition of the three solutions that fulfill the assignment constraint of the first process. The probabilities of the three solutions depend on the parameters $\theta_1$ and $\theta_2$, and the objective of the parameter optimization is to prepare one of the three solutions. In general, if the number of Edge nodes is $N$, the circuit prepares a superposition of $N+1$ solutions, which correspond to assigning the process to one of the Edge nodes or to the Cloud. Since, in general, a state is a superposition of $2^{(N+1)}$ basis states, the circuit allows exploring a sub-space with an exponentially lower number of dimensions. Of course, the same applies to the other processes and to the related assignment constraints.

The four qubits shown at the bottom of the circuit correspond to the slack variables $b_{jk}$, as described in Section \ref{SubSecProblems}. Each of these qubits undergoes an $R_Y$ rotation, so that the resulting state is a superposition, with real coefficients, of the states $\ket{0}$ and $\ket{1}$. The classical optimization routine takes the duty of preparing one of the two basis states. Overall, the objective of the circuit is to prepare a state that, beyond matching all of the constraints, minimizes the objective function.

With this ansatz, the number of tunable parameters is:

\begin{equation}
\label{eq:nParams}
    \Theta = P \cdot N + 
    \sum_{j \in \mathcal{N}}
    \lceil \log_2(B_j+1) \rceil
\end{equation}
which, for the system under consideration, is equal to $10$.
The number of two-qubit gates is $G_2 = 2 \cdot P \cdot N$ (in the example, this equals $12$), and the circuit depth is $D_2 = 2 \cdot N$ ($4$, in the example).

\subsubsection{Ansatzes A2 and A3}
\begin{figure}[!tb]
    \centering
\includegraphics[width=0.80\columnwidth, trim=0cm 0cm 0cm 0cm]{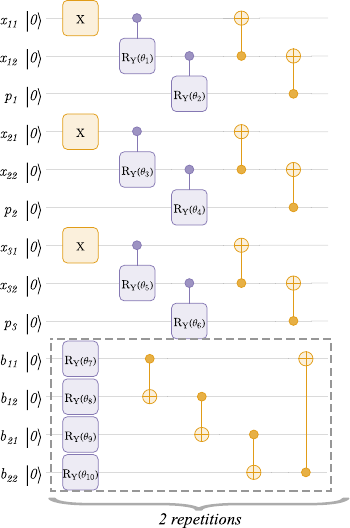}
    \caption{Ansatz A2. Besides fulfilling the assignment constraints, this ansatz creates entanglement among all the slack qubits.}
    \label{fig:ansatz_withslack}
\end{figure}

When using the Ansatz A1, the slack qubits can be prepared in any superposition of $\ket{0}$ and $\ket{1}$, with real coefficients; but these qubits are neither entangled with each other nor with the other qubits. The entanglement of the slack qubits could help to explore more regions of the Hilbert space and, possibly, avoid the exploration being trapped in local minima \cite{ParametrizedExpressibility}. Following this argument, we defined the Ansatzes A2 and A3. With the former one, shown in Figure \ref{fig:ansatz_withslack} for the sample scenario under consideration, all the slack qubits are entangled with each other, by using the \textit{TwoLocal} circuit\footnote{Please see https://qiskit.org/documentation/stubs/qiskit.circuit.library.\\TwoLocal.html}, i.e., a parameterized circuit consisting of alternating rotation layers, in this case $R_Y$ rotations, and entanglement layers, obtained with CNOT gates. The number of layer repetitions was set to 2, which is a common choice made to limit the circuit depth of the circuit and, at the same time, give enough freedom to explore the Hilbert space. The number of tunable parameters of the Ansatz A2 is larger, and equal to:

\begin{equation}
\label{eq:nParams}
    \Theta = P \cdot N + 2 \cdot \sum_{j \in \mathcal{N}} \lceil \log_2(B_j+1) \rceil
\end{equation}
which for the example under consideration is equal to $14$. The number of two-qubit gates is also larger than the one needed by the Ansatz A1. It is equal to:

\begin{equation}
\label{eq:nGates}
    G_2 = 2\cdot P \cdot N +
    2 \cdot \sum_{j \in \mathcal{N}}
    \lceil \log_2(B_j+1) \rceil
\end{equation}
giving a total of $20$ gates for our specific case. The circuit depth $D_2$ is equal to $4$.

\begin{figure}[!tb]
    \centering
    \includegraphics[width=0.8\columnwidth, height=1.2\columnwidth, trim=0cm 0cm 0cm 0cm]{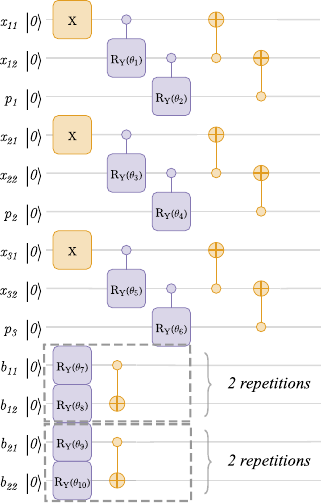}
    \caption{Ansatz A3. This ansatz creates entanglement among the slack qubits related to the same node.}
    \label{fig:ansatz_withslack_by_node}
\end{figure}

The Ansatz A3 was defined after observing that the residual capacity of a node is defined by the concurrent values assumed by the slack variables related to the same node. Therefore, the corresponding slack qubits are more entwined with each other than they are with the slack qubits of another node. In fact, Ansatz A3 creates entanglement, again using the \textit{TwoLocal} circuit, only among the slack qubits related to the same node, as shown in Figure \ref{fig:ansatz_withslack_by_node}. The number of tunable parameters of the Ansatz A3 is the same as the Ansatz A2. 
The number of two-qubit gates is lower than the Ansatz A2, as it is given by:
\begin{equation}
\label{eq:nGates}
    G_2 = P \cdot N + 2 \cdot \sum_{j \in \mathcal{N}}
    \big( \lceil \log_2(B_j+1) \rceil -1 \big)
\end{equation}
which in the case under consideration is equal to $16$. The circuit depth $D_2$ is equal to $4$, as for the Ansatz A2.

\subsubsection{Ansatz A4}

\begin{figure}[!tb]
    \centering
    \includegraphics[width=0.95\columnwidth, trim=0cm 0cm 0cm 0cm]{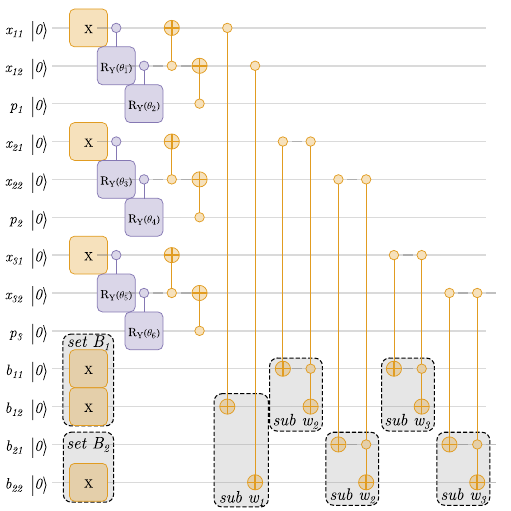}
    \caption{Ansatz A4. The state of slack qubits is determined by the values of assignment qubits. First, slack qubits are set to the capacities of respective nodes (here $B_1=3$ and $B_2=2$), then the process weights ($w_1$, $w_2$ and $w_3$) are subtracted if the related processes are assigned to the nodes.}
    \label{fig:ansatz_withslack_bounded_slack}
\end{figure}

The Ansatz A4 is defined to exploit the fact that the values of slack variables are univocally determined by the assignment of processes to nodes. Figure \ref{fig:ansatz_withslack_bounded_slack} shows a circuit that computes the correct values of the slack qubits. Notice that this computation is performed in superposition on each state defined by the assignment qubits $x_{ij}$.
With this latter ansatz, the number of tunable parameters is lower w.r.t. the others, and it is equal to $\Theta = P \cdot N$, which in our case reduces to $6$. The slack qubits represent the residual capacities of nodes; therefore, the gates on these qubits are used, first, to set the nominal capacities of the nodes, expressed as binary numbers, then to subtract from such values the weights of the processes that are assigned to the nodes. 

The exact values of $G_2$ and $D_2$ depend on the specific settings. However, it is possible to compute, for the general case, their orders of magnitude, which are polynomial in the number of processes and in the capacity of nodes:

$\mathcal{O}(G_2) = P\cdot (\sum_{j \in \mathcal{N}}(\lceil \log_2(B_j+1) \rceil)^3$      

$\mathcal{O}(D_2) = P \cdot (max_{j \in \mathcal{N}} \lceil log_2(B_j+1) \rceil)^2$

These results are derived from the work in \cite{incremental.gidney}, which shows that the subtraction of a constant can be achieved with a logarithmic number of multi-controlled-NOT gates and a logarithmic number of auxiliary qubits, and the work in \cite{PhysRevA.106.042602}, which shows that a multi-controlled-NOT gate can be obtained with a circuit topology having linear depth and a quadratic number of two-qubit gates.

The computation of the parameters can be easily generalized for the other types of problems. In particular, the presence/absence of the Cloud is taken into account by substituting the quantity $N$ with $N-1+c$, where $c$ is a binary variable that is set to 1 in presence of the Cloud and to 0 otherwise. To consider the consolidation scenario, it suffices to substitute each capacity $B_i$ with the residual capacity $\hat{B}_j$, defined in (\ref{eq:residualB}). For a quick comparison, Table \ref{tabAnsatz} reports the values of the parameters for the four considered ansatzes. 


\begin{table*}[!tb]
\caption{Comparison of the four ansatzes. It is worth noting that all the reported quantities are polynomial w.r.t. the problem size.}
\centering
\renewcommand{\arraystretch}{1.4}
\begin{tabular}{|c|c|c|c|}
\hline
\textbf{Ansatz} & \textbf{No. of tunable parameters $\Theta$} & \textbf{No. of 2-qubit gates $G_2$} & \textbf{Circuit depth $D_2$} \\ \hline
A1    & 
$ P \big(N-1+c\big) +  \sum_{j \in \mathcal{N}} \lceil \log_2(\hat{B}_j+1) \rceil $   &
$2 P \big(N-1+c\big) $ &  $2 N$\\
\hline
A2     & 
$P  \big(N-1+c\big) + 2 \sum_{j \in \mathcal{N}} \lceil \log_2(\hat{B}_j+1) \rceil$&
$2 P \big(N-1+c\big) + 2 \sum_{j \in \mathcal{N}} \lceil \log_2(\hat{B}_j+1) \rceil$ &
$2 \sum_{j \in \mathcal{N}} \lceil \log_2(\hat{B}_j+1) \rceil$\\ \hline
A3     &
$P \big(N-1+c\big) + 2  \sum_{j \in \mathcal{N}} \lceil \log_2(\hat{B}_j+1) \rceil$ &
$P \big(N-1+c\big) + 2 \sum_{j \in \mathcal{N}} \big( \lceil \log_2(\hat{B}_j+1) \rceil -1 \big)$ &
$2 \cdot \max_{j \in \mathcal{N}} \big( \lceil \log_2(\hat{B}_j+1) \rceil -1 \big)$ \\ \hline
A4     &  $P N$    &
$\mathcal{O}(P\cdot (\sum_{j \in \mathcal{N}}(\lceil \log_2(\hat{B}_j+1) \rceil)^3)$       &
$\mathcal{O}(
P \cdot (
max_{j \in \mathcal{N}} \lceil \log_2(\hat{B}_j+1) \rceil
)^2
)$ \\ \hline
\end{tabular}
\label{tabAnsatz}
\end{table*}

\section{Results}
\label{secResults}



In this section, we present a set of experimental results achieved with Variational Quantum Algorithms. \blue{The source code used to run the experiments and retrieve the results is available on a git repository\footnote{\blue{Code repository at \url{https://gitlab.com/eig-icar-cnr/quantum-edge}}}.} The objective of the experiments was to assess the performance of QAOA and VQE when solving a simple assignment problem. To this aim, we executed the algorithm using the IBM ``Qasm'' simulator, both in its standard configuration, without noise, and after applying to the simulator the noise model and the qubit coupling of real quantum hardware, specifically, \textit{ibmq\_hanoi}, an IBM quantum device of type Falcon R5, equipped with 27 qubits.
Both noiseless and noisy simulations are interesting, the former to anticipate the performance that can be achieved by real hardware in the next years, the latter to approximate the results on currently available hardware, without incurring the long delays experienced today, because these devices are made available to numerous users and institutions worldwide. We also report results obtained on real hardware, the \textit{ibmq\_hanoi} quantum computer, with a twofold goal: validate the results obtained on noisy simulators and assess the computing time experienced on quantum hardware. \blue{The optimization of the parameters, both in simulation and in the execution with quantum hardware, was performed with the Constrained Optimization by Linear Approximation (Cobyla) algorithm.}


We evaluated the following performance indices:
\begin{itemize}
    \item the \textit{success probability}, $P_{best}$, computed as the probability with which the final measurement produces the optimal solution of the problem. The number of measurements (\textit{shots}) was set to 4096 for all the experiments;
    \item the \textit{feasible solution probability}, $P_{feas}$, defined as the probability that the final measurement gives an admissible solution, i.e., a solution (optimal or non-optimal) that satisfies the constraints;
    \item the \textit{computing time} of the algorithm execution.
\end{itemize}

Moreover, because $P_{best}$ and $P_{feas}$ decrease with the size of the problem, coherently with the general idea proposed in \cite{montanezbarrera2023unbalanced}, we define the Coefficients of Performance for the success and the feasible solution probabilities, respectively  $C_{best}$  and $C_{feas}$.  They are defined as the ratio between $P_{best [feas]}$ and the probability of obtaining an optimal [feasible] solution as a random guess: 
$C_{best [feas]}=P_{best[feas]}/(N_{best[feas]}\cdot 1/2^Q)$, where $N_{best[feas]}$ is the number of optimal [feasible] solutions. These indices help us to evaluate the ability of the quantum algorithms to amplify the probabilities of measuring useful basis states.

The rest of this section discusses the most interesting results of the experiments.

\subsection{Comparing QAOA and VQE}
\label{subsecComparingQaoaVqe}

The first set of results shows that VQE performs better than QAOA, thanks to use of an ansatz that reduces the size of the search space. The comparison is made using a simple EOHL problem (see Section \ref{SubSecProblems}), in which we need to assign three processes to two Edge nodes, i.e., $P=3$, and $N=2$. The two nodes have different capacity, with $B_1=3$ and $B_2=2$, and their minimum loads are set, respectively, to $T_1=2$ and $T_2=1$.

The values and the weights are assigned as follows:

\begin{itemize}
\item $(v_{11}, v_{12}, v_{21}, v_{22}, v_{31}, v_{32}) = (2, 1, 3, 1, 2, 1) $
\item $(w_1, w_2, w_3) = (2, 1, 1) $
\end{itemize}
and, thus, the value of the penalty is $A=11$.

For this problem, the number of needed qubits is $Q=8$, see expression (\ref{eq:Q_EOHL}). 
The problem is expressed as:

\begin{align*}
     \max ~ & (2x_{11} + x_{12} + 3x_{21} + x_{22} + 2x_{31} + x_{32}) \\
    & 2 \leq 2x_{11} + x_{21} + x_{31} \leq 3 \\
    & 1 \leq 2x_{12} + x_{22} + x_{32} \leq 2
\end{align*}

After the definition of slack variables, the inclusion of the constraints into the objective function, and the variable substitutions (\ref{eq:substitutionx}), we obtain the Ising problem and then the Hamiltonian operator, expressed as:

{\footnotesize
\begin{align*}
\textbf{H} = 55.5
+12   (\textbf{Z}_{1})
-10.5 (\textbf{Z}_{2})
+7    (\textbf{Z}_{3})
-5    (\textbf{Z}_{4})\\
+6.5  (\textbf{Z}_{5})
-5    (\textbf{Z}_{6})
+5.5  (\textbf{Z}_{7})
-5.5  (\textbf{Z}_{8})\\
+5.5  (\textbf{Z}_{1}\otimes \textbf{Z}_{2})
+11   (\textbf{Z}_{1}\otimes \textbf{Z}_{3})
+11   (\textbf{Z}_{1}\otimes \textbf{Z}_{5})\\
+11   (\textbf{Z}_{1}\otimes \textbf{Z}_{7})
+11   (\textbf{Z}_{2}\otimes \textbf{Z}_{4})
+11   (\textbf{Z}_{2}\otimes \textbf{Z}_{6})\\
+11   (\textbf{Z}_{2}\otimes \textbf{Z}_{8})
+5.5  (\textbf{Z}_{3}\otimes \textbf{Z}_{4})
+5.5  (\textbf{Z}_{3}\otimes \textbf{Z}_{5})\\
+5.5  (\textbf{Z}_{3}\otimes \textbf{Z}_{7})
+5.5  (\textbf{Z}_{4}\otimes \textbf{Z}_{6})
+5.5  (\textbf{Z}_{4}\otimes \textbf{Z}_{8})\\
+5.5  (\textbf{Z}_{5}\otimes \textbf{Z}_{6})
+5.5  (\textbf{Z}_{5}\otimes \textbf{Z}_{7})
+5.5  (\textbf{Z}_{6}\otimes \textbf{Z}_{8})
\end{align*}
}

\begin{figure}[!tb]
    \centering
    \includegraphics[width=0.90\columnwidth, trim=0cm 0cm 0cm 0cm]    {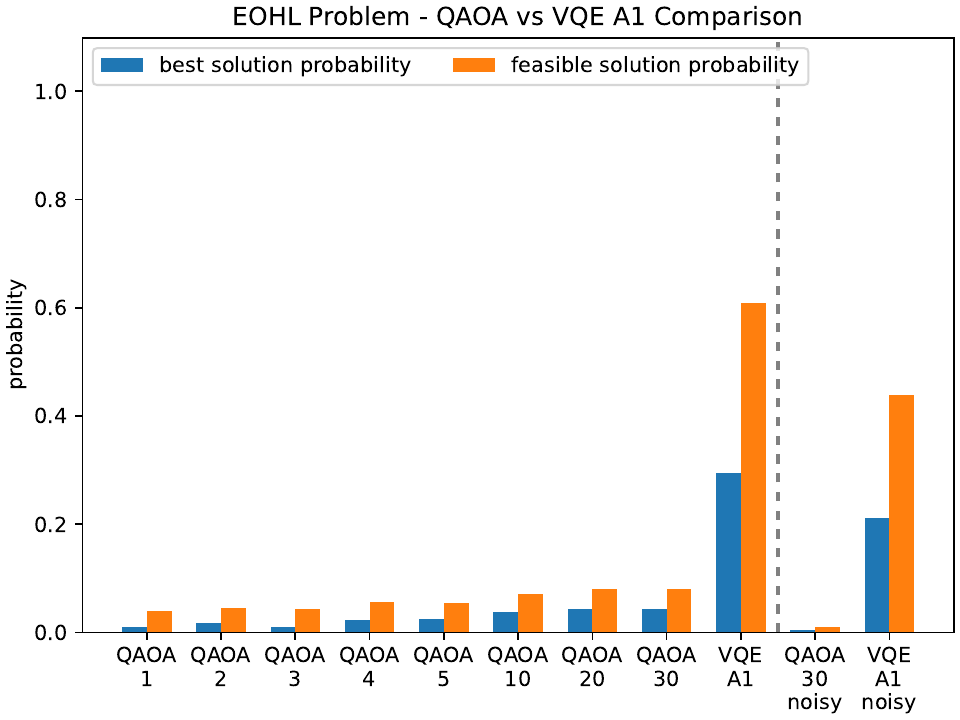}
    \caption{Probabilities $P_{best}$ and $P_{feas}$, with noiseless and noisy simulations, for an EOHL problem with 8 qubits. Comparison between QAOA with growing number of reps, and VQE equipped with the simplest ansatz, Ansatz A1.}
    \label{fig:comparison_VQE_QAOA}
\end{figure}

\begin{figure}[!tb]
    \centering
    \includegraphics[width=0.90\columnwidth, trim=0cm 0cm 0cm 0cm]    {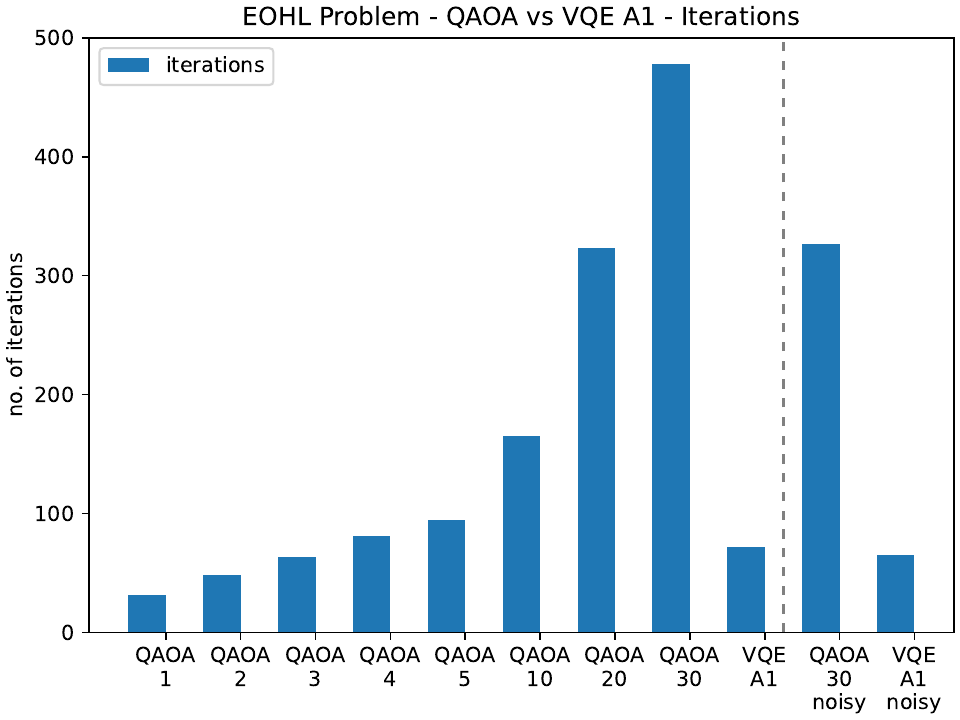}
    \caption{Average number of iterations, with noiseless and noisy simulations, for an EOHL problem with 8 qubits. Comparison between QAOA, with growing number of reps, and VQE equipped with the simplest ansatz, Ansatz A1.}
    \label{fig:comparison_VQE_QAOA_iter}
\end{figure}


\blue{Figure \ref{fig:comparison_VQE_QAOA} shows the results, in terms of probabilities $P_{best}$ and $P_{feas}$, obtained for this problem with noiseless and noisy simulation on the IBM simulator ``Qasm''. We compare the results achieved with QAOA, with a number of repetitions \textit{reps} between 1 and 30 (only the case with \textit{reps} equal to 30 is reported for noisy simulations), and with VQE equipped with the simplest ansatz, i.e., Ansatz A1. Though the theory predicts that the performances of QAOA improve with more repetitions, the results show that the improvement is slow, and much better performances are achieved with VQE. In the presence of noise, the probabilities decrease with respect to the noiseless experiments, but the performances of VQE are still much better than those of QAOA. The slow performance increase of QAOA is very likely due to the growing number of parameters, related to the different repetitions, which must be optimized. Indeed, the average number of iterations grows proportionally to the number of repetitions, as shown in Figure \ref{fig:comparison_VQE_QAOA_iter}. As also shown in a previous work, i.e., \cite{IEEESmartGrid}, noise and decoherence effects increase with the depth of the circuit, a phenomenon that impacts QAOA much more than VQE\footnote{\blue{With QAOA, the depth of the circuit is proportional both to the number of repetitions, which are not needed in VQE, and to the number of iterations, which is larger in QAOA than in VQE.}}. When combining the results of two figures, we can see that VQE is definitely preferable, since it ensures much better performances with a much shallower circuit, thanks to its ability to reduce the search space, even with the simplest ansatz.} 

\begin{figure}[!b]
    \centering
    \includegraphics[width=0.90\columnwidth, trim=0cm 0cm 0cm 0cm]    {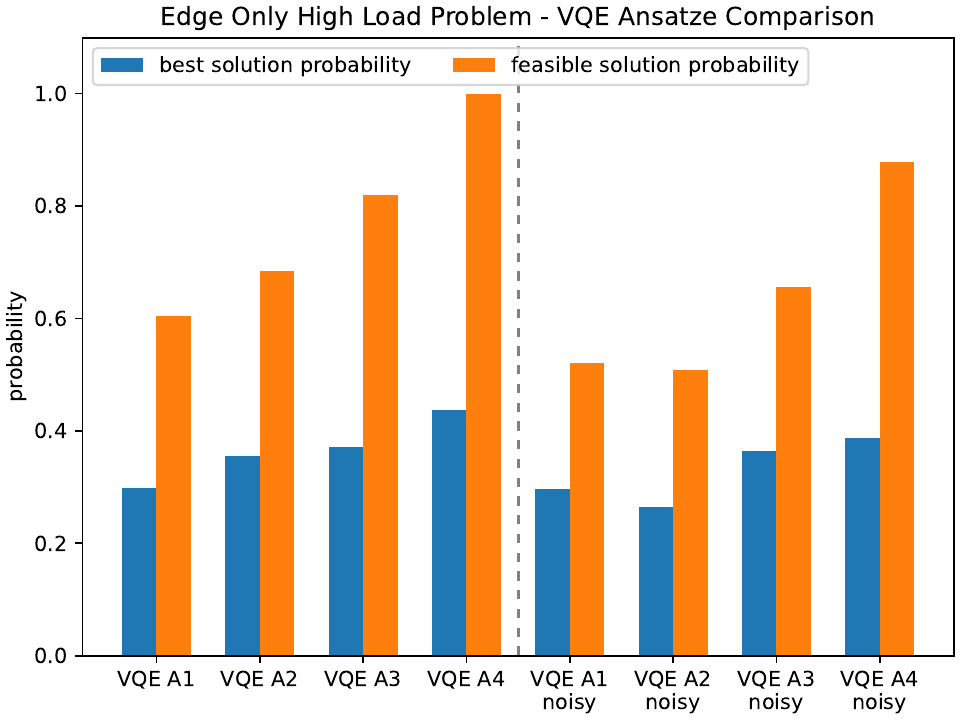}
    \caption{Probabilities $P_{best}$ and $P_{feas}$, with noiseless and noisy simulations, for an EOHL problem with 8 qubits. Comparison among the four VQE ansatzes.}
    \label{fig:comparison_VQE_Ansatze}
\end{figure}


\begin{figure}[!h]
    \centering
    \includegraphics[width=0.90\columnwidth, trim=0cm 0cm 0cm 0cm]    {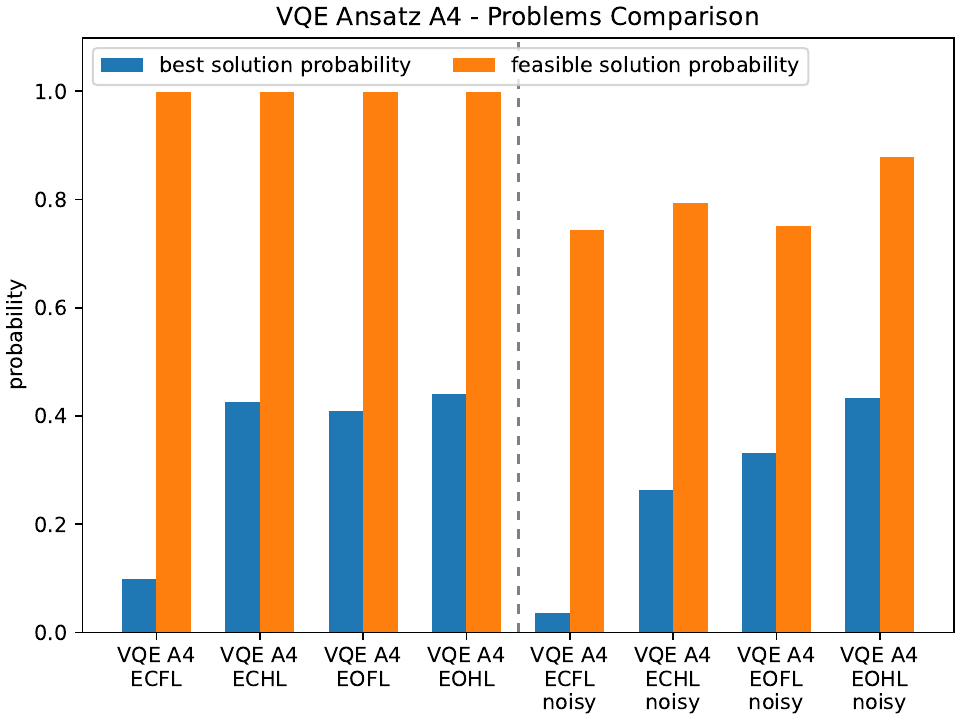}
    \caption{Probabilities $P_{best}$ and $P_{feas}$, with noiseless and noisy simulations, when using the Ansatz A4 with the four defined problems.}
    \label{fig:comparison_VQEA4_problems}
\end{figure}


\subsection{Comparing Ansatzes}
\label{subsecComparingAnsatzes}

The conclusion that VQE performs better than QAOA was confirmed by many experiments, also with different problems. The next step of the study was, then, to compare the effectiveness of the different ansatzes illustrated in Section \ref{subsecVQE}. Figure \ref{fig:comparison_VQE_Ansatze} reports the probabilities $P_{best}$ and $P_{feas}$ obtained, with noiseless and noisy simulations, when solving the same EOHL problem with 8 qubits. Both the performance indexes improve when entanglement is allowed on the slack qubits (Ansatz A2), and even more when the entanglement is restricted on the slack qubits related to the same node (Ansatz A3).
However, the performances are even better when the formulation of the problem allows for the values of the slack variables to be determined by the assignment of processes to nodes, which is achieved with the Ansatz A4. In particular, in noiseless simulations, the probability of preparing an admissible solution is almost equal to one, and almost half of the times the solution is the best one.

\subsection{Comparing Problems}
\label{subsecComparingProblems}

Since Ansatz A4 proved to be the most effective, we adopted it in simulation experiments where the objective was to solve the different problems described in Section \ref{SubSecProblems}. For all of the four problems, the number of processes and Edge nodes, and the capacities of the latter, were set as specified in Section \ref{subsecComparingQaoaVqe}: $P=3$, $N=2$, $B_1=3$ and $B_2=2$. The differences consist in the presence/absence of the Cloud and in the number of slack qubits used to determine the residual capacity, i.e., 1 for the EOHL and ECHL problems, and 2 for the EOFL and ECFL problems. This study is interesting because the four problems are formulated with different number of qubits, as summarized in Table \ref{tabProblemsQubits}, and therefore the results help to understand how the size of the problem impacts on the performances.

The results are reported in Figure \ref{fig:comparison_VQEA4_problems}: the values of $P_{best}$ and $P_{feas}$, obtained without noise, are very similar for the EOHL, EOFL and ECHL problems, showing that the different sizes have a limited impact. On the other hand, for the ECFL problem, the probability $P_{feas}$ is still high, while $P_{best}$ decreases remarkably, and is about 0.1. It must be noted, however, that the search space for this problem is very big, since the number of dimensions is equal to $2^{13}$. The size of the search problem is taken into account through the computation of the Coefficient of Performance, as discussed for the next set of results.


\begin{table}[!tb]
\caption{Number of qubits, number of best and feasible solutions, and total number of potential solutions, for the different problems, with $P$=$3$ and $N$=$2$.}
\centering
\renewcommand{\arraystretch}{1.4}
\begin{tabular}{|c|c|c|c|c|}
\hline
\textbf{Problem} & \textbf{Number of qubits  $Q$} & \textbf{\#Best} & \textbf{\#Feasible} & \textbf{\#Total} \\ \hline
EOHL    & $P \cdot N + 2 = 8$ & 2 & 4 & 256\\ \hline
EOFL    & $P \cdot N + 4 = 10$ & 2 & 4 & 1024\\ \hline
ECHL    & $P \cdot (N+1) +2 = 11$ & 2 & 6 & 2048\\ \hline
ECFL    & $P \cdot (N+1) + 4 = 13$ & 2 & 21 & 8192\\ \hline
\end{tabular}
\label{tabProblemsQubits}
\end{table}

\subsection{Execution Time and Scalability}
\label{subsecTime}
To investigate in more depth the efficiency and scalability of VQE with Ansatz A4, we run a set of experiments on simulated and real quantum hardware, considering the ECHL problem with a growing number of processes, and thus a growing number oƒ qubits. 

The real hardware exploited is the \textit{ibmq\_hanoi} quantum computer, equipped with 27 qubits.
We adopted two different procedures for noise mitigation: a default scheme (``Quantum HW'') that mitigates readout errors only, and an error mitigation layer based on Probabilistic Error Cancellation \cite{gupta2023probabilistic} (``Quantum HW mitigated''), which promises to be more effective but can be more computational intensive. The results were averaged on 5 runs.
We also executed simulations experiments on a laptop with Intel Core i71280P 2.00 GHz processor and 32 GB of RAM, again with two settings: the first (averaged on 1000 runs) without noise, the second (averaged on 100 runs) with the noise model of \textit{ibmq\_hanoi}.
Finally, for comparison purposes, we executed the renowned classical CPLEX solver, \blue{which provides the exact solution of the problem}\footnote{Documentation at https://qiskit-community.github.io/qiskit-optimization/stubs/qiskit\_optimization.algorithms.CplexOptimizer.html}.
The settings are summarized in Table \ref{tab:comparison_VQEA4_setup}.

\begin{table*}[!t]
\caption{Experimental settings for effectiveness and scalability assessment}
\centering
\renewcommand{\arraystretch}{1.4}
\begin{tabular}{|c|c|c|c|}
\hline
\textbf{Name} & \textbf{Execution Environment}  & \textbf{Noise} & \textbf{Runs} \\ \hline
Simulation No-Noise    &  Intel Core i71280P 2.00 GHz 32GB RAM & Noiseless simulation & 1000 \\ \hline
Simulation Noise &  Intel Core i71280P 2.00 GHz 32GB RAM & Simulation with \textit{ibmq\_hanoi} Noise Model & 100 \\ \hline
Quantum HW & \textit{ibmq\_hanoi} Quantum Computer & Default readout errors mitigation only & 5 \\ \hline
Quantum HW mitigated & \textit{ibmq\_hanoi} Quantum Computer & Probabilistic Error Cancellation & 5 \\ \hline
Classical & Intel Core i71280P 2.00 GHz 32GB RAM & Not Applicable & 100 \\ \hline
\end{tabular}
\label{tab:comparison_VQEA4_setup}
\end{table*}

\begin{figure}[!tb]
    \centering
    \includegraphics[width=0.90\columnwidth, trim=0cm 0cm 0cm 0cm] 
    {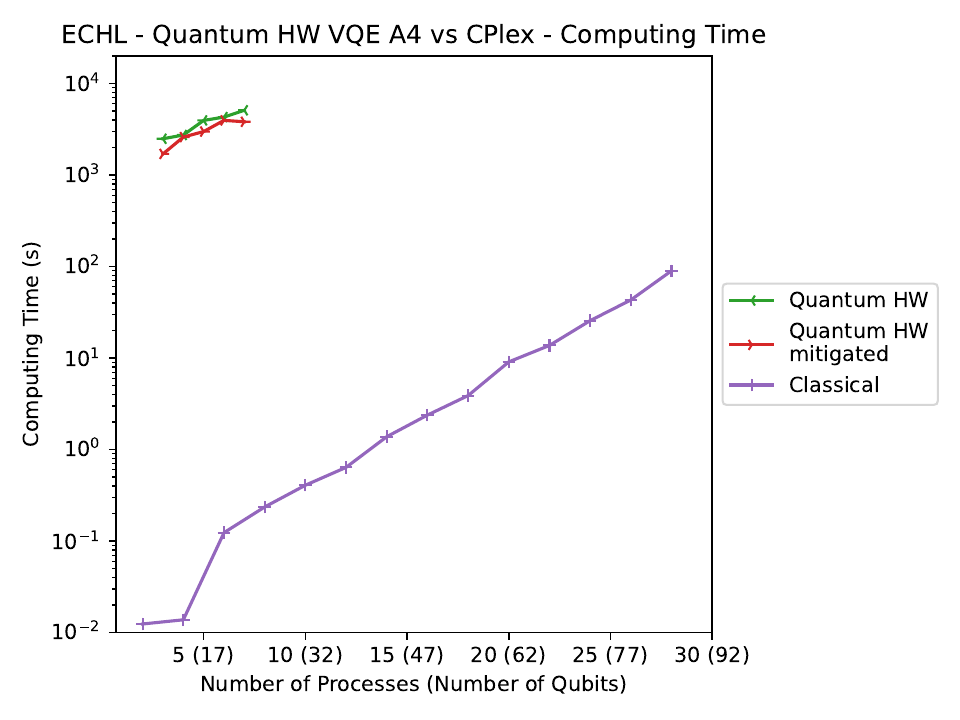}
    \caption{Comparison between VQE (Ansatz A4) on real quantum hardware and classical CPLEX algorithm in solving the allocation problem with a growing number of processes. }
    \label{fig:comparison_VQE-Quantum-vs-CPLEX}
\end{figure}

\begin{figure}[!tb]
    \centering
    \includegraphics[width=0.90\columnwidth, trim=0cm 0cm 0cm 0cm]   {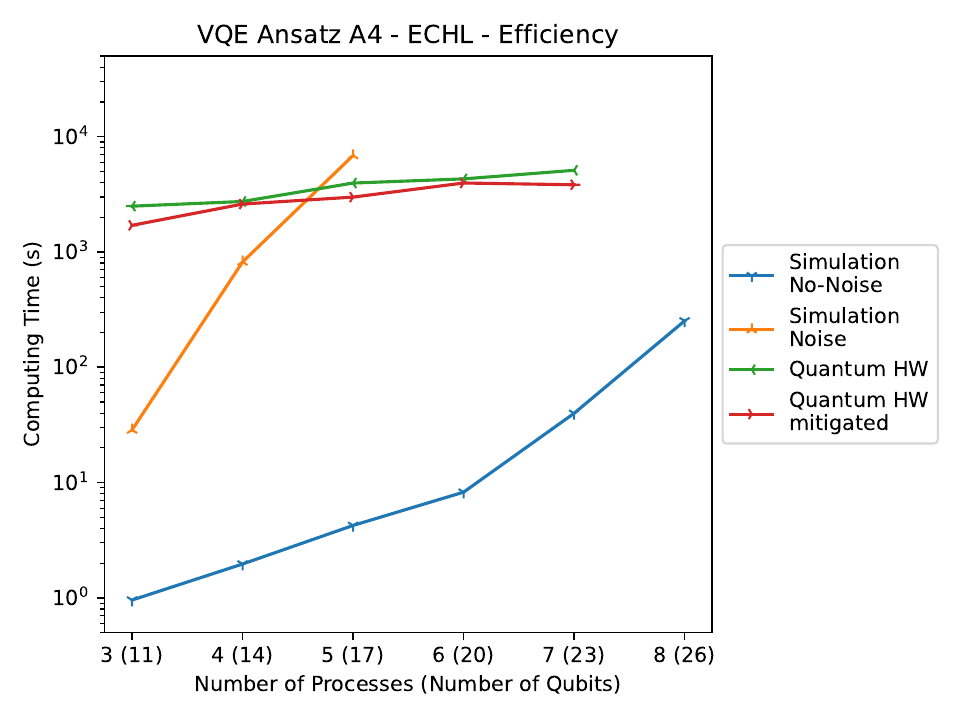}
    \caption{Computing times when using VQE and Ansatz A4, for the ECHL problem with two nodes and growing number of processes.}
    \label{fig:comparison_VQEA4_ECHL_efficency}
\end{figure}

\begin{figure}[!tb]
    \centering
    \includegraphics[width=0.90\columnwidth, trim=0cm 0cm 0cm 0cm]    {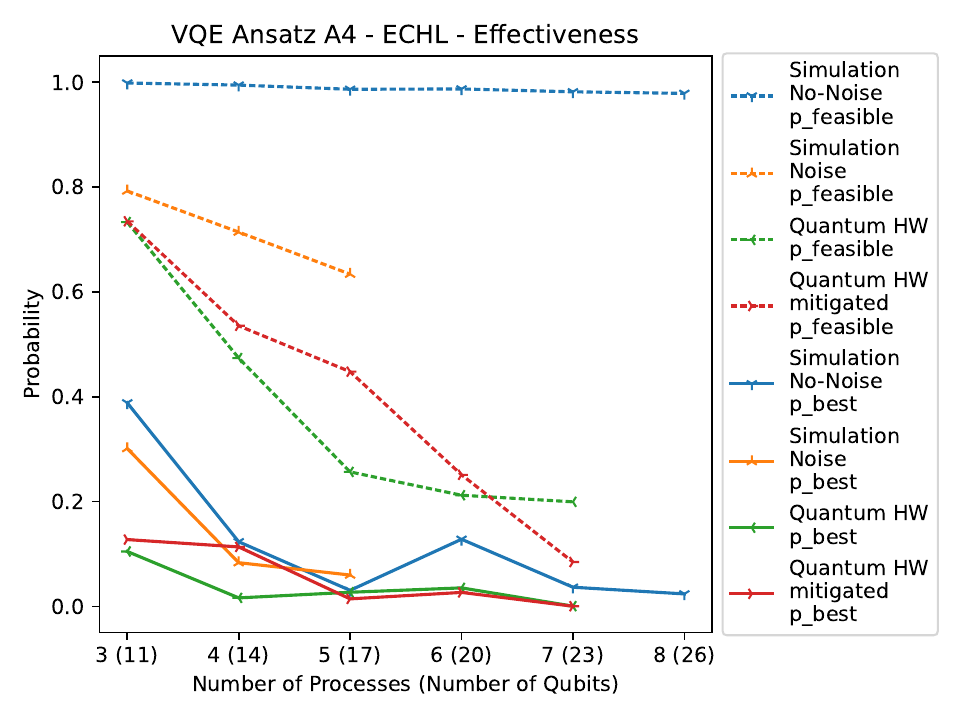}
    \caption{Probabilities $P_{best}$ and $P_{feas}$ when using VQE and Ansatz A4, for the ECHL problem with two nodes and growing number of processes.}  \label{fig:comparison_VQEA4_ECHL_Effectiveness}
\end{figure}

In Figure \ref{fig:comparison_VQE-Quantum-vs-CPLEX}, we compare the computing times required by quantum hardware to those of the CPLEX solver, for a growing number of processes. A notable distinction in the scalable behavior of the two approaches emerges: the time required by CPLEX exhibits exponential growth vs. the number of binary variables, whereas the quantum computing time experiences a moderate increase rate. Though the computing times of the quantum algorithm are much longer for the small-size problems for which direct comparison is possible,
the discrepancy in growth rates instills optimism regarding the potential for a substantial acceleration of quantum computing algorithms for solving large-scale problems, as soon as enhancements in quantum hardware make them capable of tackling such problems. It is worth noting that ``Quantum HW mitigated'' is faster than ``Quantum HW'': the reason is that, even if the probabilistic error cancellation technique requires more computational time for each iteration,
the optimizer requires fewer iterations to converge.

\begin{figure}[!tb]
    \centering
    \includegraphics[width=0.90\columnwidth, trim=0cm 0cm 0cm 0cm]    {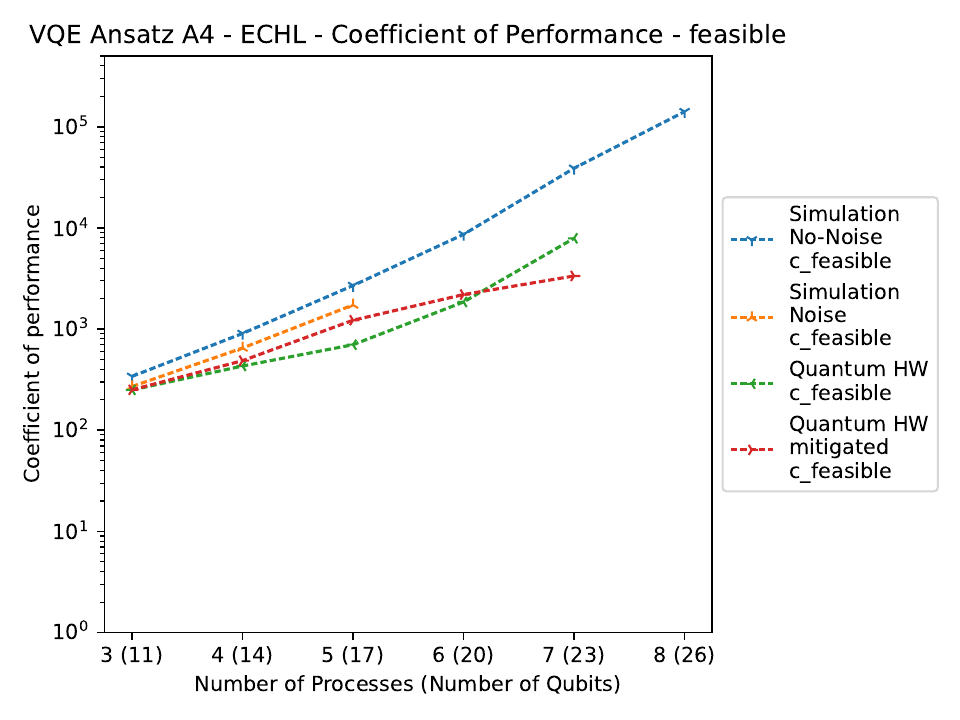}
    \caption{Coefficient of Performance when using VQE and Ansatz A4, for the ECHL problem with two nodes and growing number of processes.}
    \label{fig:comparison_VQEA4_ECHL_coef_perf}
\end{figure}

Figure \ref{fig:comparison_VQEA4_ECHL_efficency} compares the computing times recorded when executing the quantum algorithm on real hardware and on simulators. \blue{The simulation settings are faster than the execution on the real device, for small number of qubits;
but, as expected, they exhibit an exponential trend, due to the exponential increase in the dimensions of the Hilbert space of the quantum register, which
confirms that quantum simulation is not applicable for large problems}\footnote{The ``Simulation Noise'' setting overtakes all the other trends when reaching 17 qubits, and thus no further result is provided.}.
%
Figure \ref{fig:comparison_VQEA4_ECHL_Effectiveness} reports the probabilities $P_{best}$ and $P_{feas}$ obtained with the different settings. As expected, the best results are those of the ``Simulation No-Noise'' setting, as it corresponds to executing VQE on an ideal quantum computer. The trends exhibited by ``Quantum HW'' and ``Quantum HW mitigated'' are reasonably comparable, considering that each result is averaged over 5 runs. Both $P_{best}$ and $P_{feas}$ decrease when the number of qubits grows (and, thus, the search space grows exponentially), but $P_{feas}$ remains above 10\% also when considering 23 qubits.
Finally, Figure \ref{fig:comparison_VQEA4_ECHL_coef_perf} compares the Coefficients of Performance of the same experimental settings. It is worth noting that this index appears to grow exponentially w.r.t. the number of processes (and thus the size of the solution space), which is a promising result for the application of the proposed VQE ansatz on even larger problems, when better quantum hardware will be broadly available in the current NISQ era.

\section{Conclusions and Future Work}
\label{secConclusions}

In this paper, we investigated the possibility of exploiting Quantum Computing to solve the process assignment problem in the Edge/Cloud continuum, i.e., choosing where (Edge or Cloud) to assign a set of processes. Specifically, after the problem is translated into an Ising optimization task, we showed it being tackled by employing two different Variational Quantum Algorithms, i.e., QAOA and VQE.

Our first aim was, indeed, to compare these two approaches; in particular, we exploited the freedom of defining customized and problem-inspired ansatzes in the VQE approach, in order to experiment with different strategies in solving a simple version of the problem. In particular, we explored the VQE algorithm with four different ansatzes, aimed at assessing different entanglement patterns in the quantum register, and compared its performances with those of the QAOA approach, mainly in terms of success probability and execution time. 

We found that VQE performs better than QAOA. Indeed, despite the latter being known to converge with increasing the number of repetitions, it turns out that this convergence is very slow. On the other hand, a suitably customized VQE ansatz is able to provide better performances with a much smaller number of repetitions, due to the fact that the search space (i.e., the set of states accessible by the parameterized unitary) is limited.
This is particularly relevant as, when performing experiments with real devices, the latter are prone to noise and decoherence errors, so that the effective performances decrease in a substantial way as the number of repetitions increases. 

When adopting VQE, the structure of the variational circuit becomes crucial: it needs to be expressible enough to contain the solution of the problem; but, at the same time, the space of accessible states must be restricted in order to give rise to an affordable optimization procedure. For the Edge/Cloud assignment problem, we showed that what matters is the entanglement between the qubits associated to slack variables and those associated to the corresponding nodes. Other possible entanglement patterns, while always helping improve the results with respect to the case in which slack qubits are factorized, have a less pronounced effect. This can be understood intuitively: as each slack variable helps saturating an inequality concerning the load of a specific Edge node, the corresponding qubits need to be entangled.
We also analyzed resource scaling when increasing the size of the problem.
Despite the limiting factors of present-day devices, the results show that the scaling trend of the quantum algorithm, experienced on real hardware, is better than the classical one, which, as expected, increases exponentially.

\blue{Some interesting avenues for future work are: (i) as bigger and less error-prone hardware becomes available, perform more experiments to improve the comparison between classical and quantum algorithms; (ii) continue the investigation of ansatzes and search for a possible general strategy to adapt them to the specific problems and the input data embedded in the quantum circuit; (iii) use quantum algorithms to tackle optimization and assignment problems regarding the management of the Internet of Things, among which: physical and virtual resource management \cite{IOT-ResMan}, network deployment and sensor placement \cite{IoT-SensorPlacement}, and battery-driven energy optimization \cite{IoT-EnergyMgt}; (iv) explore the possibility of distributing the execution of Variational Quantum Algorithms, an approach that can bring opportunities but also hard challenges, related not only to the distribution of the computation per se but also to the availability of multiple quantum computers and to the necessity of coordinating them.}

\section*{Acknowledgments}
\footnotesize{
This work was partially funded by ICSC – Italian Research Center on High Performance Computing, Big Data and Quantum Computing, funded by European Union – NextGenerationEU, PUN: B93C22000620006. This work was also supported by European Union - NextGenerationEU - National Recovery and Resilience Plan (Piano Nazionale di Ripresa e Resilienza, PNRR) - Project: “SoBigData.it - Strengthening the Italian
RI for Social Mining and Big Data Analytics” - Prot. IR0000013 - Avviso n. 3264 del 28/12/2021, and by European Union - NextGenerationEU - the Italian Ministry of University and Research, PRIN 2022 “INSIDER: INtelligent ServIce Deployment for advanced cloud-Edge integRation”, grant n. 2022WWSCRR, CUP H53D23003670006. JS acknowledges the contribution from PRIN (Progetti di Rilevante Interesse Nazionale) TURBIMECS - Turbulence in Mediterranean cyclonic events, grant n. 2022S3RSCT CUP H53D23001630006, CUP Master B53D23007500006. 
}

\bibliographystyle{IEEEtran}
\bibliography{bibliography}

\newpage

\section*{Biography Section}

 


\vspace{-33pt}

\begin{IEEEbiography}[{\includegraphics[width=1in,height=1.25in,clip,keepaspectratio]{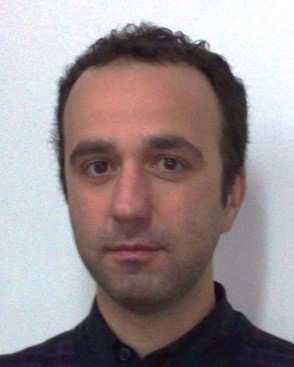}}]{Carlo Mastroianni}
received the Laurea degree and the PhD degree in computer engineering from the University of Calabria, Italy, in 1995 and 1999, respectively.
He is a Director of Research with ICAR-CNR, Rende, Italy. He has coauthored over 100 papers published in international journals and conference proceedings. His current research interests include distributed computing, internet of things, cloud computing, bio-inspired algorithms, smart grids and quantum computing.
\end{IEEEbiography}

\begin{IEEEbiography}[{\includegraphics[width=1in,height=1.25in,clip,keepaspectratio]{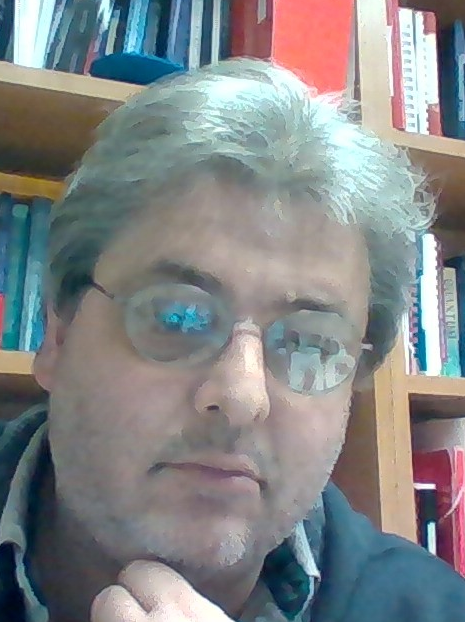}}]{Francesco Plastina} received his Ph.D. in Physics at University of Calabria in 2000, where he is now full professor in theoretical condensed matter physics. He is coauthor of more than 80 research papers, mainly focusing on quantum information theory, quantum thermodynamics, quantum coherence and correlations, and open quantum systems.  
\end{IEEEbiography}


\begin{IEEEbiography}[{\includegraphics[width=1in,height=1.25in,clip,keepaspectratio]{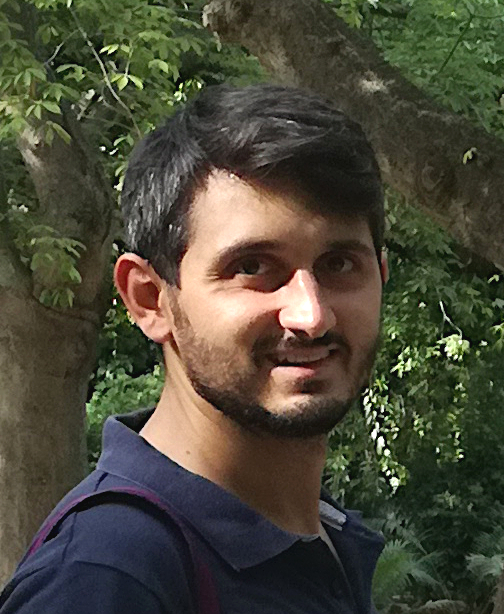}}]{Jacopo Settino}
received his master's degree in Physics at the University of Calabria, Italy, in 2015.
He got his PhD Doctor Europaeus title in Physical, Chemical and Materials Sciences and Technologies, in 2019.
After a two-year post-doc at SPIN-CNR he is a researcher at the Physics Department of University of  Calabria, Rende. His many research activities run on quantum algorithms, topological superconductors, disordered and interacting quantum systems.
\end{IEEEbiography}

\begin{IEEEbiography}[{\includegraphics[width=1in,clip,keepaspectratio]{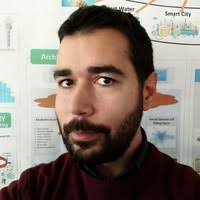}}]{Andrea Vinci}
received a PhD in system engineering and computer science from the University of Calabria, Rende, Italy. He is a Researcher with ICAR-CNR, Rende, Italy,
since 2012. His research mainly focuses on the Internet of Things, cyber-physical systems and Quantum Computing. He has authored or co-authored researches on the definitions of platforms and methodologies for the design and implementation of cyber-physical systems, and on distributed algorithms for the efficient control of urban and building infrastructures based on artificial and swarm intelligence.
\end{IEEEbiography}



\vfill

\EOD

\appendix
\section{Calculation of the coefficients of the Ising Model}

In this Appendix, we show how the coefficients in the QUBO expression for the ECFL problem of Eq. (\ref{eq:extObj}) are related to those of the Ising problem in Eq. (\ref{eq:IsingHamiltonian}), Section \ref{subsecStepsToIsing}.
Starting from Eq. (\ref{eq:extObj}), that we report below for clarity,   
\begin{multline*}
     \min \bigg( - \sum_{i \in \mathcal{P},j \in \mathcal{N}}v_{ij}x_{ij}  + \\
      A \cdot \sum_{j \in \mathcal{N}}\Big(
        B_j - \sum_{i \in \mathcal{P}} w_ix_{ij} + \sum_{k=1}^{\lceil \log_2(B_j+1) \rceil} 2^{(k-1)} b_{jk}
      \Big)^2 + \\
      A \cdot \sum_{i \in \mathcal{P}}\Big(
      1 - \sum_{j \in \mathcal{N}} x_{ij} + p_i\Big)^2     
      \bigg) ,
\end{multline*}
we define new binary variables \(z_{ij}\), \(z_{b_{jk}}\), and \(z_{p_{i}}\) as transformations of the original variables \(x_{ij}\), \(b_{jk}\), and \(p_{i}\) to accommodate a formulation that aligns with the Ising model. Specifically, these new variables are defined such that 
\begin{equation*} 
x_{ij} = \frac{1-z_{ij}}{2}, \quad b_{jk} = \frac{1-z_{b_{jk}}}{2}, \quad \text{and} \quad p_{i} = \frac{1-z_{p_{i}}}{2}.
\end{equation*}
This transformation ensures that the original binary nature of \(x_{ij}\), \(b_{jk}\), and \(p_{i}\) is preserved in the new variables \(z_{ij}\), \(z_{b_{jk}}\), and \(z_{p_{i}}\), which now assume values of \(+1\) or \(-1\). Following this, we expand the squares, eliminate all constant terms, group terms with identical binary variables, and introduce Kronecker deltas to enable a comprehensive summation. Consequently, we derive the following ILP formulation of the ECFL problem:
\begin{multline*}
\min \biggl\{  \sum_{i \in \mathcal{P}, j \in \mathcal{N}} \bigl[ \frac{1}{2} v_{ij} + \frac{A}{2} ( c_3 w_i+1-N) \bigr] z_{ij}  \\
- \frac{A}{2} c_3 \sum_{j \in \mathcal{N}} \sum_{k=1}^{\lceil \log_2(B_j+1) \rceil} 2^{(k-1)} z_{b_{jk}}  + \frac{A}{2} (1-N) \sum_{i \in \mathcal{P}} z_{p_{i}} \\
+ \frac{A}{2} \sum_{\substack{i,i' \in \mathcal{P} \\ i' < i}} \sum_{\substack{j,j' \in \mathcal{N} }}  w_i w_{i'} \delta_{j,j'} z_{ij} z_{i' j'} \\
+ \frac{A}{2} \sum_{\substack{j,j' \in \mathcal{N} }} \sum_{\substack{k,k'=1 \\ k' < k}}^{\lceil \log_2(B_j+1) \rceil} 2^{(k + k' - 2)} \delta_{j,j'} z_{b_{jk}} z_{b_{j' k'}}  \\
+ \frac{A}{2} \sum_{i \in \mathcal{P}} \sum_{\substack{j,j' \in \mathcal{N} }} \sum_{k=1}^{\lceil \log_2(B_j+1) \rceil} 2^{(k-1)} w_i \delta_{j,j'} z_{ij} z_{b_{j' k}} \\
+ \frac{A}{2} \sum_{\substack{i,i' \in \mathcal{P}}} \sum_{\substack{j,j' \in \mathcal{N} \\ j' < j}} \delta_{i,i'} z_{i j} z_{i' j'} 
+ \frac{A}{2} \sum_{\substack{i,i' \in \mathcal{P} }} \sum_{j \in \mathcal{N}} 
\delta_{i,i'} z_{ij} z_{p_{i'}} \biggl\},
\end{multline*}
with \[c_4 = c_3^2 + \sum_{i \in \mathcal{P}} w_i^2 + \sum_{k=1}^{\lceil \log_2(B_j+1) \rceil} 2^{2(k-1)}   \]
and \[ c_3 = 2B_j - \sum_{i \in \mathcal{P}} w_i + \sum_{k=1}^{\lceil \log_2(B_j+1) \rceil} 2^{(k-1)}. \] 
The optimization problem is reformulated into an Ising model, where the original variables and their interactions are encoded into a flat index space. The objective function becomes:
\begin{equation}
\min \bigg( \sum_{\alpha=1}^{Q} h_\alpha \cdot z_\alpha - \sum_{\alpha=1}^{Q}\sum_{\beta=1}^{\alpha-1} J_{\alpha \beta} \cdot z_\alpha \cdot z_\beta \bigg)
\end{equation}
In this representation, \(h_\alpha\) encapsulates coefficients for linear terms, directly correlating with the effects of individual variables, while \(J_{\alpha \beta}\) captures the coefficients for quadratic terms, describing the interactions between couples of variables. 
The equivalence between the two expressions is established through the following relationships among the coefficients.

\textbf{Coefficients} $h_{\alpha}$:

\begin{itemize}
    \item For \(\alpha \in I_{P N}= \{ 1, \ldots, P \times N \} \), corresponding to pairs \(\{i, j\}\) from sets \(\mathcal{P}\) and \(\mathcal{N}\):
    \[ h_{\alpha} = \frac{1}{2} v_{ij} + \frac{A}{2} (c_3 w_i + 1 - N). \]
    
    \item For \(\alpha \in I_{b}= \{ P \times N + 1, \ldots, P \times N + \sum_{j} \log_2(B_j+1) \} \), corresponding to \(b_{jk}\):
    \[ h_{\alpha} = -\frac{A}{2} c_3 2^{(k-1)}. \]
    
    \item For \(\alpha \in I_{p}= \{ P \times N + \sum_{j} \log_2(B_j+1) + 1, \ldots, Q \} \), corresponding to \(p_i\), where \(Q = P \times (N+1) + \sum_{j} \log_2(B_j+1)\):
    \[ h_{\alpha} = \frac{A}{2} (1-N). \]
\end{itemize}

\textbf{Interaction Coefficients} $J_{\alpha\beta}$:
\begin{itemize}
    \item For $\alpha, \beta \in I_{P N}$, corresponding to $\{i, j\}$ and $\{i', j'\}$ with $i' < i$:
    \[ J_{\alpha\beta} = \frac{A}{2} w_i w_{i'} \delta_{jj'} + \frac{A}{2} \delta_{ii'}, \]
    where $\delta_{jj'}$ is the Kronecker delta, ensuring $j = j'$.
    
    \item For $\alpha, \beta \in I_{b}$, corresponding to $b_{jk}$ and $b_{j'k'}$ with $k' < k$:
    \[ J_{\alpha\beta} = \frac{A}{2} 2^{(k + k' - 2)} \delta_{jj'}. \]
    
    \item For $\alpha \in I_{P N}$ and$\beta \in I_{b}$, corresponding to $\{i, j\}$ and $b_{j'k}$:
    \[ J_{\alpha\beta} = \frac{A}{2} 2^{(k-1)} w_i \delta_{jj'}. \]
    
    \item For $\alpha, \beta \in I_{p}$, corresponding to $p_i$ and $p_{i'}$:
    \[ J_{\alpha\beta} = \frac{A}{2} \delta_{ii'}. \] 
\end{itemize}

\end{document}